%% file: paperARIS2.tex
\documentclass[a4paper,11pt]{article}
\pdfoutput=1 % if your are submitting a pdflatex (i.e. if you have
\usepackage{jinstpub} % for details on the use of the package, please
\usepackage{lineno}
\title{\boldmath A title with some math: $x=1$}

\usepackage{makecell}
\usepackage{todonotes}
\usepackage{float}

\usepackage[separate-uncertainty,retain-explicit-plus,per-mode=symbol,binary-units]{siunitx}
\usepackage{array,mathtools,dcolumn}
\usepackage{amsmath}
\usepackage[below]{placeins}
\usepackage{tikz}
\usepackage{afterpage}
\usepackage{paralist}
\usepackage{subfig}
\usepackage{graphicx}
\usepackage{listings}
\usepackage{array}
\usepackage{comment}
\usepackage[version=4]{mhchem}
\usepackage{multirow}
\usepackage{eurosym}
\usepackage{pagecolor}
\usepackage{blindtext}
\usepackage{enumitem}
\usepackage{array}
\usepackage{xspace}
\newcolumntype{L}[1]{>{\raggedright\let\newline\\\arraybackslash\hspace{0pt}}m{#1}}
\newcolumntype{C}[1]{>{\centering\let\newline\\\arraybackslash\hspace{0pt}}m{#1}}
\newcolumntype{R}[1]{>{\raggedleft\let\newline\\\arraybackslash\hspace{0pt}}m{#1}}

\newcommand{\am}{$\mathrm{^{241}Am}$\xspace}
\newcommand{\ba}{$\mathrm{^{133}Ba}$\xspace}

\title{Characterization of the scintillation time response  of liquid argon detectors for dark matter search}

\newcommand{\APC}{APC, Universit\'e de Paris, CNRS, Astroparticule et Cosmologie, Paris F-75013, France}

\newcommand{\AQLNGS}{Laboratori Nazionali del Gran Sasso, Assergi AQ 67010, Italy}
\newcommand{\Houston}{Department of Physics, University of Houston, Houston, TX 77204, USA}

\newcommand{\LPNHE}{LPNHE Paris, Sorbonne Universit\'e, Universit\'e Paris Diderot, CNRS/IN2P3, Paris 75252, France}

\newcommand{\NAINFN}{Istituto Nazionale di Fisica Nucleare, Sezione di Napoli, Napoli 80126, Italy}
\newcommand{\NAUni}{Department of Physics, Universit\`a degli Studi Federico II, Napoli 80126, Italy}
\newcommand{\Princeton}{Department of Physics, Princeton University, Princeton, NJ 08544, USA}
\newcommand{\Temple}{Department of Physics, Temple University, Philadelphia, PA 19122, USA}
\newcommand{\Davis}{Department of Physics, University of California, Davis, CA 95616, USA}

\newcommand{\UCLA}{Department of Physics and Astronomy, University of California, Los Angeles, CA 90095, USA}
\newcommand{\IPNO}{Univerist\'e Paris-Saclay, IJCLab, CNRS/IN2P3, F-91405 Orsay, France}
\newcommand{\GSSI}{Gran Sasso Science Institute, L'Aquila 67100, Italy}
\newcommand{\RUL}{Royal Holloway, University of London, Department of Physics, Egham, TW20 0EX, UK}
\newcommand{\Sapienza}{Physics Department, Sapienza Universit\`a di Roma, Roma 00185, Italy}

\author[a,b]{P.~Agnes}%\affiliation{\Houston}\affiliation{\APC}
\author[c]{S.~De~Cecco}%\affiliation{\LPNHE}
\author[d,1]{A.~Fan\note{Present address: SLAC National Accelerator Laboratory, Menlo Park, CA 94025-7015, USA}}%\affiliation{\UCLA}
\author[e,f]{G.~Fiorillo}%\affiliation{\NAUni}\affiliation{\NAINFN}
\author[b]{D.~Franco}%\email{davide.franco@apc.in2p3.fr}\affiliation{\APC}
\author[g,h]{C.~Galbiati}%\affiliation{\Princeton}
\author[i]{C.~Giganti}%\affiliation{\LPNHE}
\author[l,p]{G.~Korga}%\affiliation{\Houston}\affiliation{\AQLNGS}
\author[m]{M.~Lebois}%\affiliation{\IPNO}
\author[l]{A.~Mandarano}%\affiliation{\GSSI}\affiliation{\AQLNGS}
\author[n]{C.~J.~Martoff}%\affiliation{\Temple}
\author[o]{L.~Pagani}%\email{davide.franco@apc.in2p3.fr}%\affiliation{\Davis}
\author[o]{E.~Pantic}%\affiliation{\Davis}
\author[l]{A.~Razeto}%\affiliation{\AQLNGS}
\author[a]{A.~L.~Renshaw}%\affiliation{\Houston}
\author[b,2]{Q.~Riffard\note{Present address: Lawrence Berkeley National Laboratory (LBNL), Berkeley, CA 94720-8099, USA}}%\email{riffard@apc.in2p3.fr}\affiliation{\APC}
\author[o]{B. Schlitzer}%\affiliation{\Davis}
\author[b]{A.~Tonazzo}%\affiliation{\APC}
\author[d]{H.~Wang}%\affiliation{\UCLA}
\author[m]{J.~N.~Wilson}%\affiliation{\IPNO}

\affiliation[a]{\Houston}
\affiliation[b]{\APC}
\affiliation[c]{\Sapienza}
\affiliation[d]{\UCLA}
\affiliation[e]{\NAUni}
\affiliation[f]{\NAINFN}
\affiliation[g]{\Princeton}
\affiliation[h]{\GSSI}
\affiliation[i]{\LPNHE}
\affiliation[l]{\AQLNGS}
\affiliation[m]{\IPNO}
\affiliation[n]{\Temple}
\affiliation[o]{\Davis}
\affiliation[p]{\RUL}

\abstract{
The  scintillation time response of liquid argon  has a key role in the  discrimination of electronic backgrounds in  dark matter search experiments.  However, its extraordinary rejection power can be affected by various detector effects such as the delayed light emission of TetraPhenyl Butadiene, the most commonly used wavelength shifter, and the electric drift field applied in Time Projection Chambers. In this work, we characterized the TetraPhenyl Butadiene delayed response and the dependence of the  pulse shape discrimination on the electric field, exploiting the data acquired with the ARIS, a small-scale single-phase liquid argon detector exposed to monochromatic neutron and gamma sources at the ALTO facility of IJC Lab in Orsay.}

\date{\today}

\emailAdd{davide.franco@apc.in2p3.fr}
\emailAdd{lpagani@ucdavis.edu}
\begin{document}
\maketitle
\flushbottom

%\title{Measurement of the  liquid argon time response to nuclear and electronic recoils}

%%%%%%%%%%%%%%%%%%%%

%\input{ds.def}
%\input{frontpage}

%\pacs{}
%\keywords{}

%%%%%%%%%%%%%%%%%%%%%%%%
\section{Introduction}
\label{sec:intro}
\input{intro}

\section{Experimental setup and data selection}
\label{sec:setup}
\input{setup}

\section{Characterization of the TPB time response}
\label{sec:tpb}
\input{tpb}

%\section{Pulse shape discrimination parameter $f_{p}$ dependence on electric field}
\section{Dependence of the liquid argon time response on the electric field}
\label{sec:f90}
\input{psd}

%\section{Modeling the  $f_{p}$  distribution}
%\subsection{Analytical \textit{f$_{90}$} model}
%\label{sec:f90}
%\input{f90}

%\subsection{ \textit{f$_{90}$} model validation on simulated data}
%\label{sec:f90sim}
%\input{f90sim}

%\section{$f_{90}$ dependence on the electric field}
%\label{sec:EF}
%\input{f90aris}

\section{Conclusions}
In this work, we have characterized, with the ARIS setup, two effects that impact the time response of LAr TPCs, critical for future dark matter search experiments: the fluorescence of TPB, one of the most widely used wavelength shifters, and the effect of the electric field on the LAr scintillation singlet-to-triplet ratio.  The first result confirms what has already been observed in the literature~\cite{Segreto:2014aia}, namely the presence of at least one TPB  decay component of the order of a few tens of nanoseconds.  The second measurement confirms the dependence of the LAr time response to NRs on the electric field, already observed by SCENE~\cite{Cao_2015}.  We additionally showed with ARIS data the dependence of the response to ERs, never observed before, with a trend opposite to NRs. We finally  exclude, from the multi-waveform fit,  that the electric field affects the LAr triplet de-excitation time.  

%Further investigation is needed to understand the interplay between the LAr scintillation time response and the electric field. 

\begin{acknowledgments}
%This report is based upon work supported by  the UnivEarthS LabEx program (Grants No. ANR-10-LABX-0023 and No. ANR-18-IDEX-0001).
This report is based upon work supported by the UnivEarthS LabEx program (Grants No. ANR-10-LABX-0023 and No. ANR-18-IDEX-0001) and by the National Science Foundation (Grants No. PHY-1314501, No. PHY-1314483, No. PHY-1314507, and No. PHY-1455351) and the France-Berkeley Fund (2016-0053).

%\appendix
%%\section{The derivation of the waveform model}
%%\label{sec:wfmodel}
%%\input{wfmodel}
%%\label{app:wf}

%\section{The derivation of the $f_p$ model}
%\label{sec:fpmodel}
%\input{fpmodel}
%\label{app:fp}

\end{acknowledgments}
\bibliographystyle{JHEP}
%\bibliography{bibtex/pixel2018_himac}
\bibliography{biblio}
\end{document}

%% file: intro.tex
% !TEX root = paperARIS2.tex

Recent results from DarkSide-50~\cite{DarkSide:2018fwg, Agnes:2018ves, Agnes:2018oej} and DEAP-3600~\cite{Amaudruz:2017ekt,DEAP:2019yzn} have demonstrated the large potential of liquid argon (LAr) technology in the direct search of Weakly Interacting Massive Particles (WIMPs), the leading dark matter candidate.  LAr is characterized by an extraordinary scintillation pulse shape discrimination (PSD), able to suppress the electronic recoil  background.  Such a rejection power originates from the dependence of LAr scintillation pulse on the nature of the recoil, electronic (ER) or nuclear (NR), with distinctive probabilities in populating singlet and triplet excited states.  These are characterized by de-excitation times which differ by more than two orders of magnitude, namely $\sim$6\,ns and $\sim$1600\,ns, respectively.  DEAP-3600~\cite{DEAP:2019yzn} probed such a rejection power  by a factor larger than 3$\times$10$^{7}$ in the 44--89~keV$_{er}$ energy range  by  operating the detector with atmospheric argon,  highly contaminated   in  cosmogenic $^{39}$Ar, a  $\beta$-decay isotope with a specific activity at the level of $\sim$1\,Bq/kg. 

In 2015, the DarkSide Collaboration~\cite{Agnes:2015ftt}   demonstrated that $^{39}$Ar contamination can be reduced by  a factor $\sim$1400, by extracting argon from deep underground, naturally shielded against cosmic rays.  Although underground argon allows for the relaxation of requirements on the PSD power, the ambition of the next generations of LAr  detectors~\cite{Aalseth:2017fik}  is  to  run with  exposures  equivalent to 100-1000\,ton year in a  background-free mode. This  imposes a PSD rejection power larger than 10$^9$, ideally achievable in LAr thanks to the large difference (about a factor 3) between the NR and ER probability to populate singlet or triplet states. However, light detection in LAr detectors can be delayed by the wavelength shifter, which absorbs 128~nm scintillation  photons and re-emits visible photons, detectable  by photosensors.  Several works~\cite{Chepel:2012sj,PhysRevC.78.035801,PhysRevB.27.5279,Segreto:2014aia} observed time delayed  components attributed to TetraPhenyl Butadiene  (1,1,4,4-tetraphenyl-1,3-butadiene, C$_{28}$H$_{22}$, abbreviated TPB)~\cite{instruments5010004}, a popular  wavelength shifter  with an extremely high conversion efficiency~\cite{Segreto:2014aia},  and used  by DarkSide-50 and DEAP-3600,

Another effect, potentially  impacting  the LAr PSD, is the scintillation  dependence on the electric field, reported by the SCENE Collaboration~\cite{Cao_2015}, and recently also observed  with a dual-phase Time Projection Chamber (TPC) at CERN \cite{Aimard:2020qqa}. The latter analysis  attributes such a dependence to the LAr slow scintillation component, which decreases as the drift field increases. 

In this work, we characterize the TPB fluorescence (Section \ref{sec:tpb}) and the dependence of the LAr scintillation on the electric field (Section \ref{sec:f90}), exploiting   data from ARIS, a small-scale LAr TPC exposed in 2016 to neutron and gamma beams  from the LICORNE source at the ALTO facility in Orsay.

%% file: setup.tex
% !TEX root = paperARIS2.tex

The ARIS detector has an active mass of $\sim$0.5\,kg of LAr, housed in a 7.6\,cm diameter, 7\,cm height PTFE cylinder, equipped with one 3-inch Hamamtsu R11065 photomultiplier tube (PMT) at the bottom   and seven 1-inch Hamamtsu R8520 PMTs at the top.  All the inner surfaces of the chamber are coated with evaporated TPB.  The PTFE sleeve supports a set of   copper rings connected by resistors in series to maintain a uniform electric field throughout the active argon volume.  The light yield measured at field off is equal to 6.35$\pm$0.05\,pe/keV, obtained with \am\ and \ba\ electronic recoil calibration sources.   A complete description of the chamber design and its performance can be found in Ref.~\cite{Agnes:2018mvl}. 

The LICORNE source~\cite{Lebois:2014uwk} exploits the inverse $^7$Li(p, n)$^7$Be reaction, which guarantees a quasi-monoenergetic ($\sim$1.5\,MeV), pulsed neutron beam with high collimation. In addition, 478\,keV $\gamma$s from $^7$Li* de-excitation are emitted in coincidence with the beam pulse, characterized by a period of 400~ns  and a width of 2~ns.

\begin{figure}[tb]
\centering
\includegraphics[width=0.6\columnwidth]{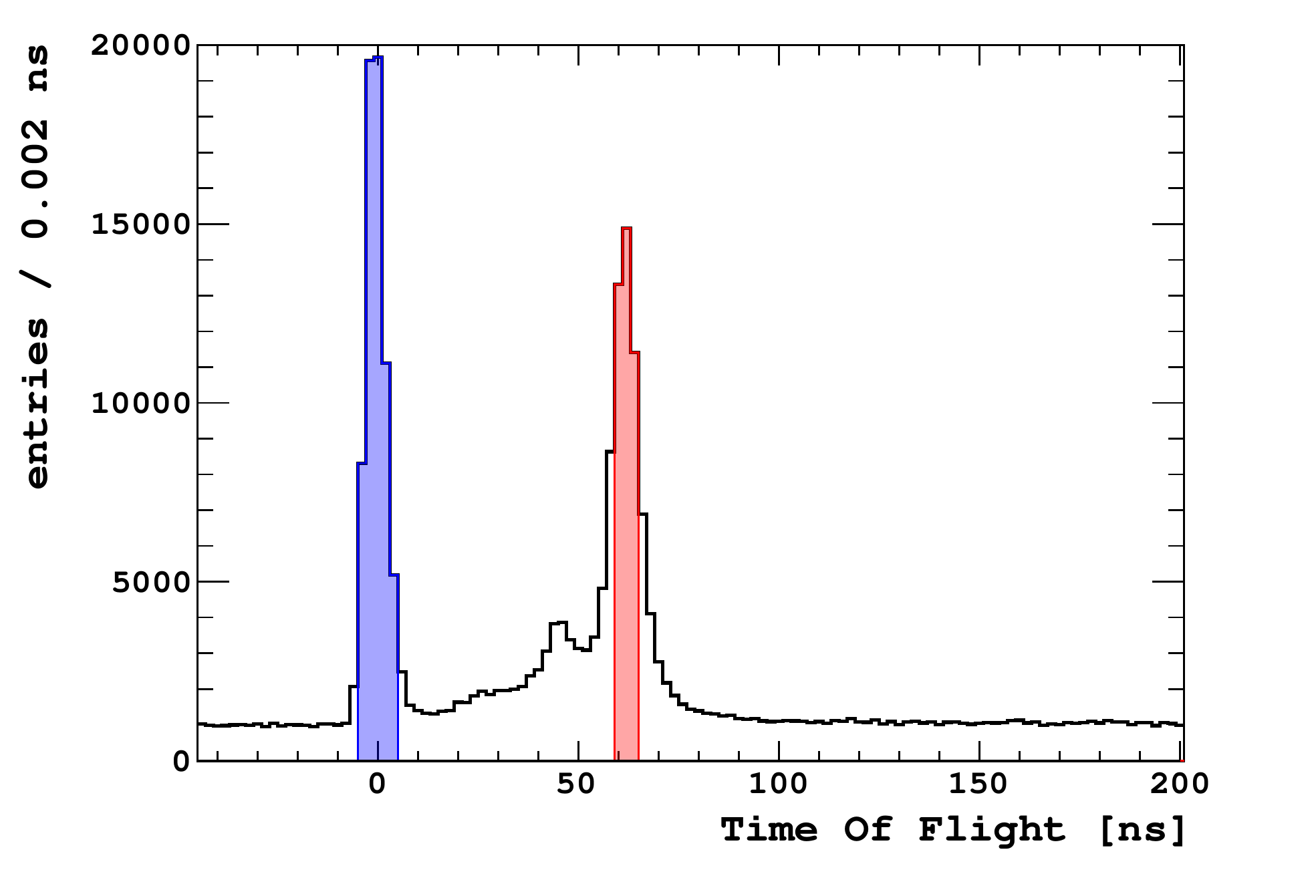}
\caption{Time-of-Flight (ToF) distribution of neutrons and $\gamma$s detected by the LAr chamber with respect to the  beam pulse. The data selection of NRs (ERs) is defined  the red (blue) ToF range.}
\label{fig:tpctof}
\end{figure}

Data have been acquired   varying the  electric drift fields from $0$ up to $500$\,V/cm. ERs, induced by 478\,keV $\gamma$s, and NRs, induced by neutrons, are efficiently separated by  looking at the time-of-flight (ToF), the difference in time between the LAr signal and the beam pulse. Their ToFs differ by $\sim$60\,ns, as shown in figure~\ref{fig:tpctof}, corresponding to  the 1\,m distance between the chamber and the source. The resolution of the ToF is measured at 1.8\,ns, dominated by the beam pulse width (1.5\,ns) and by the determination of the start time of the LAr scintillation pulse, digitized at 250\,MHz.

ARIS  is surrounded by  eight  NE213 liquid scintillator detectors from the EDEN array~\cite{CAVALLARO201365}, which tag scattered neutrons and gammas, in order to  kinematically constrain the recoil energy in the chamber, as described in Ref.~\cite{Agnes:2018mvl}. However,  in this work, we have  analyzed  dedicated runs  acquired with the beam-chamber  coincidence only, which guarantees high statistical samples of NRs (ERs), with continuous spectra up to $\sim$150\,keV$_{nr}$ ($\sim$50\,keV$_{er}$), without affecting the characterization of the time response,  expressed as a function of the number of photoelectrons (S1). High statistics  ER samples  from the \ba calibration source are also included in this study. 

\begin{figure}[tb]
\centering
\includegraphics[width=0.45\columnwidth]{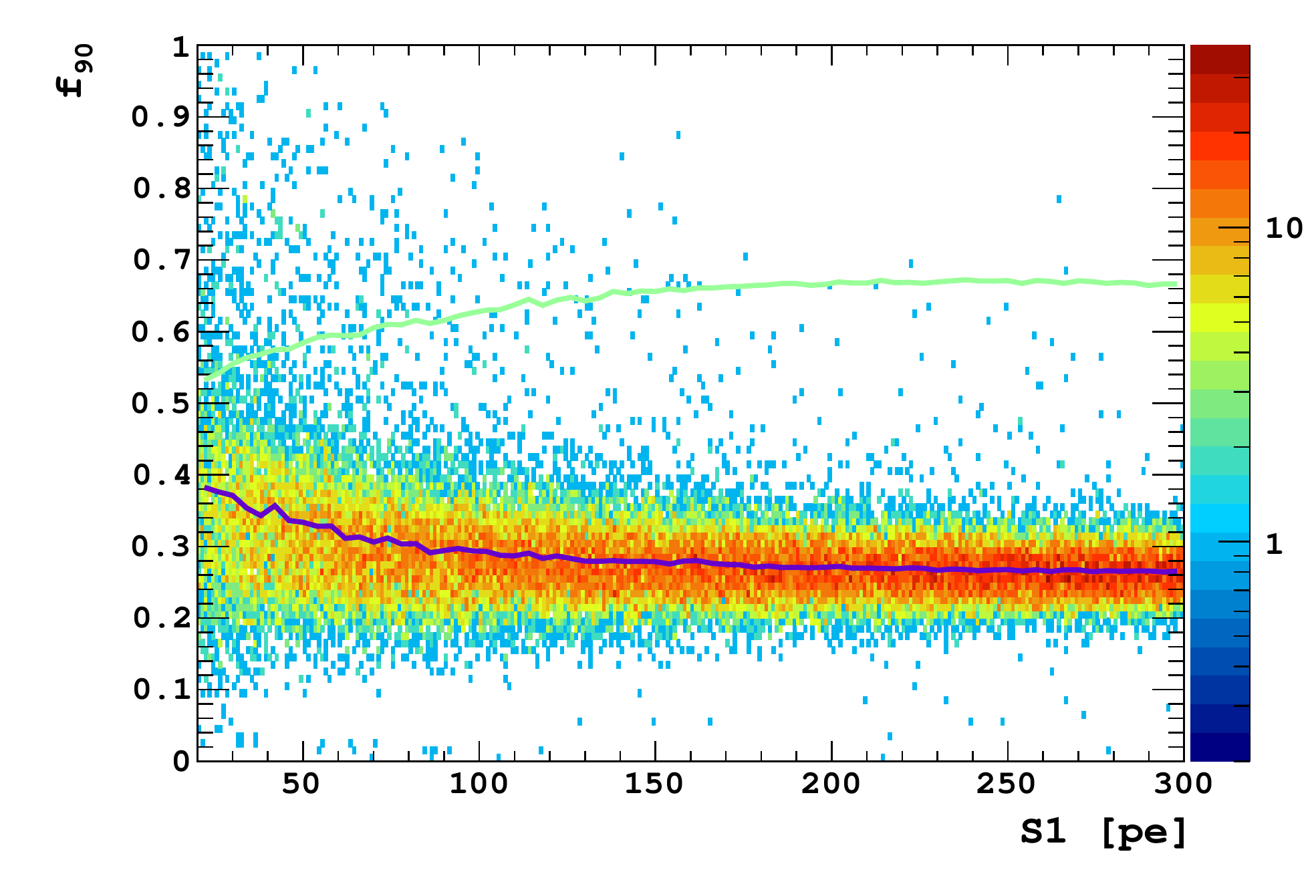}
\includegraphics[width=0.45\columnwidth]{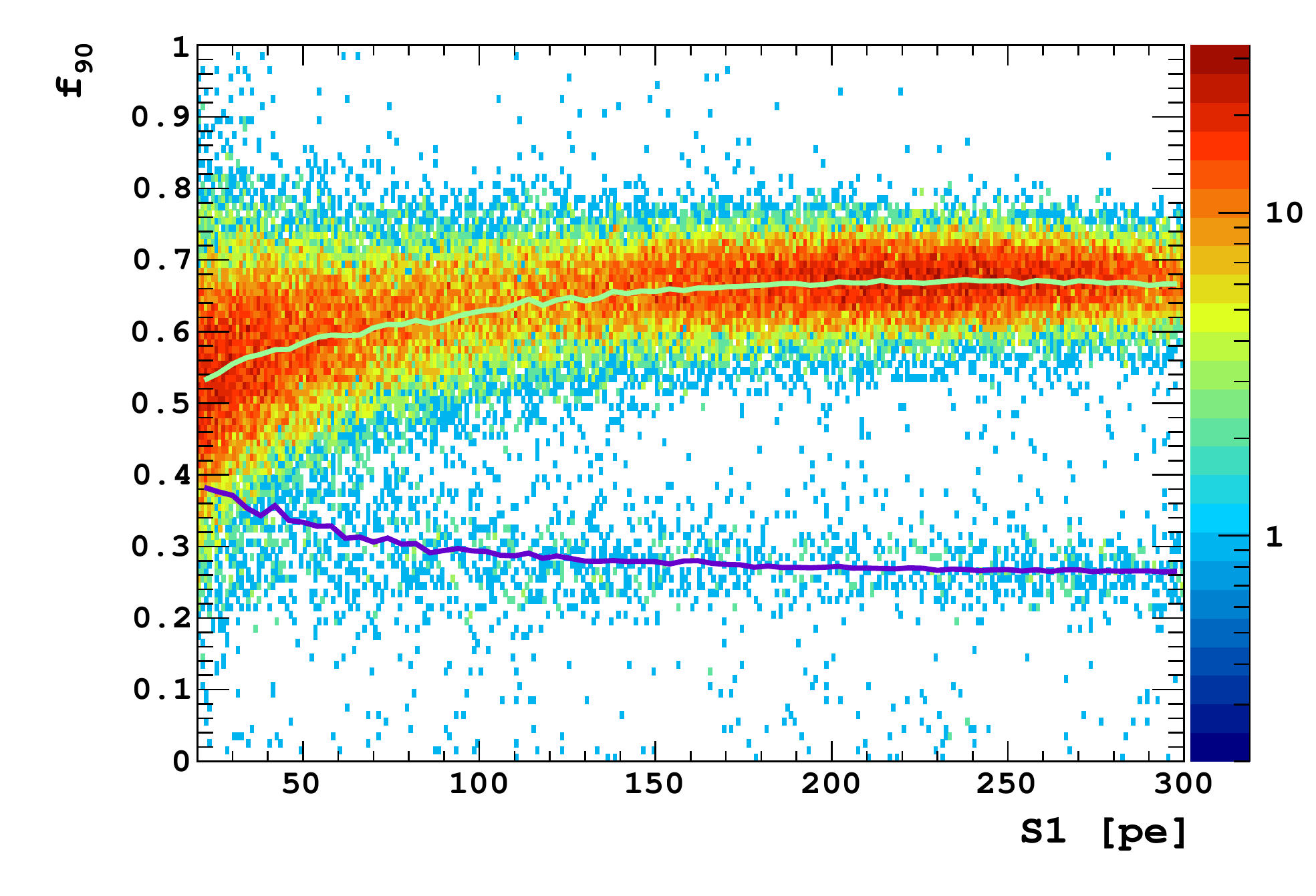}
\caption{PSD estimator, f$_{90}$,  as a function of S1 for ERs (left) and NRs (right) as selected by the ToF cuts shown in Figure \ref{fig:tpctof}. The blue and green lines correspond to the mean values of the ER and NR distributions, respectively. }
\label{fig:f90_s1}
\end{figure}

NRs and ERs are selected by requiring ToF between the beam and the TPC in the [59, 65]\,ns and [-5,5]\,ns ranges, respectively, as shown in figure~\ref{fig:tpctof}. The purity of NR sample is estimated at 97\%, with a small contamination from random coincidences, while NR contamination in  ER samples is negligible.  Such purity can be appreciated by looking at $f_{90}$, the pulse shape estimator defined  as the fraction of light observed in the first 90\,ns,  distributed around $\sim$0.3 ($\sim$0.7) for ERs  (NRs), as shown in Figure \ref{fig:f90_s1}.

%% file: tpb.tex
% !TEX root = paperARIS2.tex

Although the TPB re-emission is mostly prompt ($<$10 ns),  a residual delayed  component can affect the LAr  PSD  by lengthening photon collection times and reducing the time resolution.  The dominant TPB fluorescent component was measured by E.~Segreto~\cite{Segreto:2014aia} to be 49$\pm$1\,ns (30$\pm$1\% probability) by directly irradiating a TPB in vacuum with $\alpha$ and $\beta$ particles at room temperature.  Such a delay is comparable to the size of the prompt window used to calculate the fraction of collected  light,  $f_p$, the standard PSD estimator. As an example, DarkSide-50 uses a 90\,ns window~\cite{DarkSide:2018fwg} while DEAP-3600 a 60\,ns one~\cite{DEAP:2019yzn}, defined to fully contain photon from the singlet de-excitation. The difference between the two prompt values used by the experiments depends on the detector size and hence on the propagation length in the LAr volume. Furthermore,  E.~Segreto~\cite{Segreto:2014aia}  reported  two  additional TPB delayed components at 309$\pm$10\,ns and 3550$\pm$500\,ns with probabilities 2$\pm1$\,\%, and 8$\pm1$\,\%, respectively.

In this work, we characterize the TPB time response by directly fitting ARIS waveforms   acquired in LAr by the bottom 3-inches PMT, which collects $\sim$60\% of the light.  Top PMTs are excluded to avoid any possible distortion from averaging the waveforms of channels with different gains.  The analysis strategy is based on the simultaneous fit of multiple ER and NR waveforms acquired in different energy regimes (from 80 to 290\,pe in bins of 30\,pe). This approach allows the breaking of the degeneracy between argon scintillation and TPB parameters, since the latter do not depend on the particle nature and energy. TPB, argon scintillation, and detector parameters are therefore constrained in simultaneous fits among different samples. Conversely, the singlet-to-triplet ratio depends on the particle nature and energy, and it is included in the model as an independent parameter for each waveform.  The same procedure is repeated for 7 different electric fields (from 0 to 500\,V/cm) to probe a possible dependence of the slow argon decay component on the applied electric field, as suggested in \cite{Aimard:2020qqa}.

%The analysis strategy is based on the simultaneous fit of multiple ER and NR waveforms acquired in different energy regimes (from 80 to 290\,pe in bins of 30\,pe). The same procedure is repeated for 7 different electric fields (from 0 to 500\,V/cm).  This approach allows to break the degeneracy between argon scintillation and TPB parameters, since the latter do not depend on the particle nature and energy.   TPB, argon scintillation, and detector parameters are therefore constrained in simultaneous fits among different samples.  Conversely, the singlet-to-triplet ratio depends on the particle nature and energy, and it is included in the model as an independent parameter for each waveform. This approach allows to probe a possible dependence of the slow argon decay component on the applied electric field, as suggested in \cite{Aimard:2020qqa}. 
%The dependence of the singlet-to-triplet ratio on the electric field  will be extensively discussed in the following Section.

Waveforms are averaged after subtracting the baseline over approximately 10$^3$ events for each energy region and field, corresponding to a minimum statistics of $\sim$10$^5$ photoelectrons (pe). The error associated to each bin is defined with respect to the photon statistics.  To remove spurious events from environmental background, soft cuts are applied  on the PSD parameter, requiring $f_{90}$$<$0.5 and $f_{90}$$>$0.4, both with S1$>$80\,pe, for ERs  and NRs, respectively. The inefficiency associated to these cuts is estimated  to be  negligible,  as can be appreciated from the Figure~\ref{fig:f90_s1}. 

\subsection{The waveform model}

Waveforms are analytically modeled as the convolution of three components, namely the argon scintillation time profile, the TPB re-emission, and the detector response, which accounts for the photon propagation and the PMT jitter. 

The argon scintillation time profile is described by 
\begin{equation}
F(t, \tau_s, \tau_t, p_s) = \frac{p_s}{\tau_s} e^{-\frac{t}{\tau_s}} + \frac{1 - p_s}{\tau_t} e^{-\frac{t}{\tau_t}}, 
\end{equation}
\noindent where $\tau_s$  ($\tau_t$) is the singlet  (triplet) decay times, and  $p_s$ (1-$p_s$) the probability of populating the singlet (triplet) state.

To simplify the formalism, the TPB re-emission  is here modelled with only two  delayed components with $\tau_j$ and $p_j$ ($j$=\{1, 2\}) the decay time and intensity, respectively,   

%.  $\tau_j$ having an intensity $p_j$, being $j$=\{1, 2\}:
\begin{equation}
H(t, \tau_1, \tau_2, p_0, p_1, p_2) = p_0 + \sum_{j=1}^{2}\frac{p_j}{\tau_j} e^{-\frac{t}{\tau_j}},
\label{eqn:tpb}
\end{equation}
\noindent and where 
\begin{equation}
p_0 = 1 - p_1 - p_2, 
\end{equation}
\noindent represents the fast re-emission component, assumed instantaneous with respect to the  detector time resolution. If necessary, the model can easily be extended to more than two components. 
%For simplicity of notation,  we assume  only one TPB delayed component ($N$=1, $p^{*}$=$p_1^{*}$ and  $\tau^{*}$=$\tau_1^{*}$) in the  here-below description of the model. 

The photon propagation and the detector response are jointly described with a normal distribution, $G(t,\sigma)$, where the resolution $\sigma$, is time-independent.  

The convolution of the three components
\begin{equation}
R  =  F  \otimes H  \otimes G  
\label{eqn:conv}
\end{equation}
\noindent is computed  analytically by exploiting the associative property. 

The term representing the time response with instantaneous TPB emission is obtained by convolving the scintillation time response for each excited state with $G(t,\sigma)$, 
\begin{eqnarray}
\label{eqn:p0}
P_0^i(t,\tau_i,\sigma) & =  & \frac{e^{-\frac{t}{\tau_i}}}{\tau_i}  \otimes   \frac{e^{  {-\frac{t^2}{2 \sigma^2}} } }{\sqrt{2 \pi \sigma^2}}  \\ 
& = &\frac {1}{2 \tau_i}  \left(1 +  \text{erf}\left(\frac{t'}{\sqrt{2} \sigma}\right)\right)  e^{-t'/\tau_i}, \nonumber 
\end{eqnarray}
\noindent where 
\begin{equation}
t' =  t - \frac{\sigma^2}{\tau_i},
\label{eqn:t'}
\end{equation}
\noindent and  $i$ = \{s, t\} is referred to the singlet and triplet states.

In the presence of a delayed TPB emission, the response function is derived by  convolving for each scintillation term eq.~\ref{eqn:p0} with the TPB response 
\begin{eqnarray}
\label{eqn:p1}
P_1^{ij}(t,\tau_i, \tau_j,\sigma) & =  &  \frac{e^{-\frac{t}{\tau_i}}}{\tau_i}  \otimes    \frac{ e^{-\frac{t}{\tau_j}}}{\tau_j}   \otimes   \frac{e^{  {-\frac{t^2}{2 \sigma^2}} } }{\sqrt{2 \pi \sigma^2}}  \\ 
& = & \frac{\tau_i   P_0^i(t, \tau_i, \sigma) -  \tau_j P_0^j(t, \tau_j, \sigma)}{\tau_i - \tau_j},  \nonumber 
\end{eqnarray}
\noindent where $j$=\{1,2\} represents the two delayed TPB components. 

The overall response function, described by eq.~\ref{eqn:conv}, is the sum of different contributions from eq.~\ref{eqn:p0} and \ref{eqn:p1}

\begin{eqnarray}
R(t, \Theta) &=&   \sum_{i=\{s,t\}} (1 - p_1 - p_2) \, p_i \, P_0^i(t, \tau_i, \sigma) \nonumber \\
&   & +  \sum_{i=\{s,t\}}  \sum_{j=\{1,2\}} p_j \, p_i \,  P_1^{ij}(t, \tau_i, \tau_j,  \sigma). 
\label{eqn:model}
\end{eqnarray}
\noindent where $\Theta$ is the set of parameters including  $p_s$ and  $p_t$,  the probabilities to populate  singlet and triplet states, so that $p_s$+$p_t$=1. 

The waveform  model requires three additional parameters in order to fit  the ARIS data: the pulse amplitude ($A$), the  offset corresponding to the pulse start time ($t_0$), and noise and effects related to the baseline subtraction modeled with a constant ($C$). 
The final model derived from eq.~\ref{eqn:model} is expressed by
\begin{equation}
P(t, \Theta, t_0, A, C) =  A \times R(t - t_0, \Theta)  + C. 
\label{eqn:model2}
\end{equation}

\begin{figure}[h]
\captionsetup[subfigure]{labelformat=empty}
\centering
\includegraphics[width=0.31\textwidth]{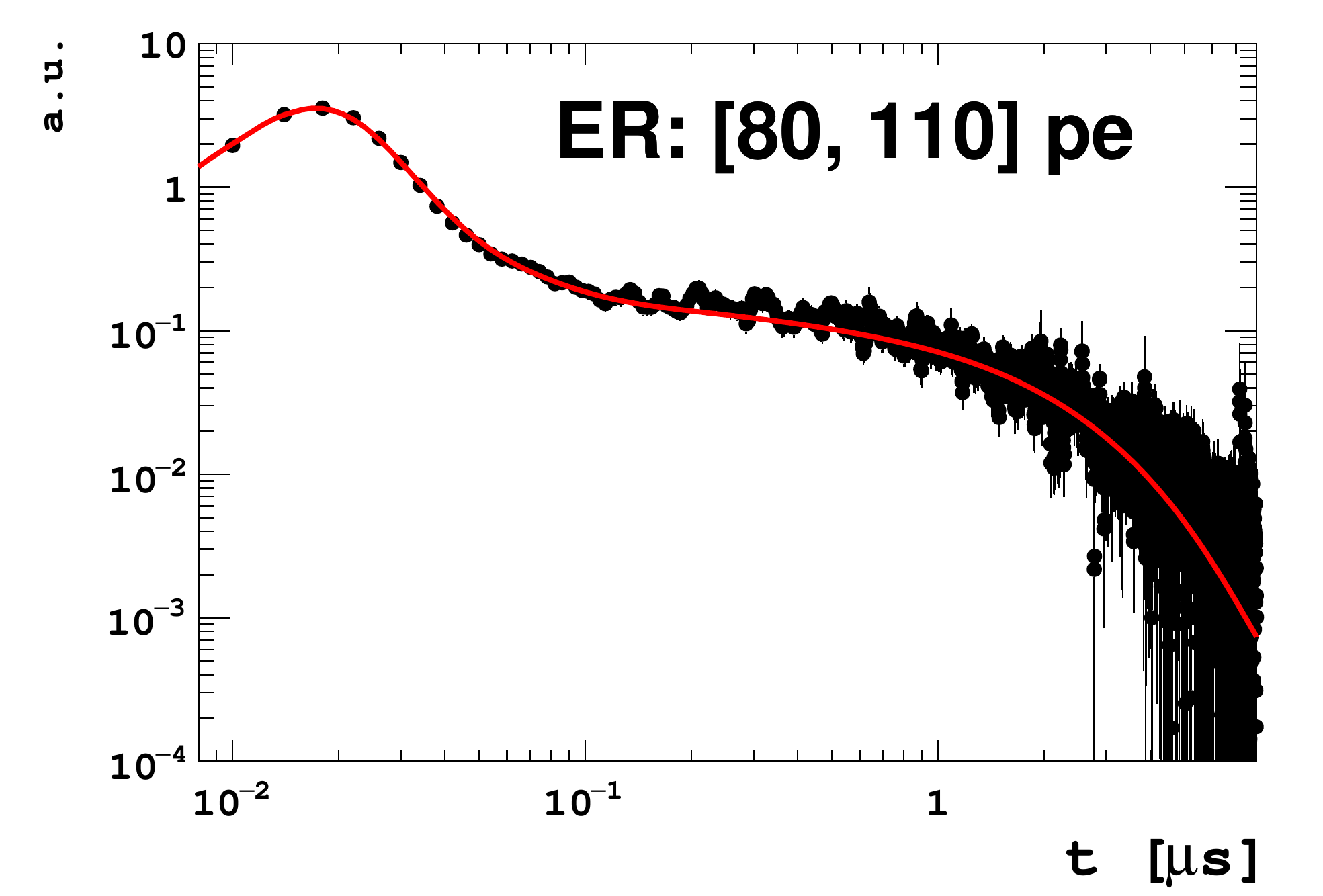} 
\includegraphics[width=0.31\textwidth]{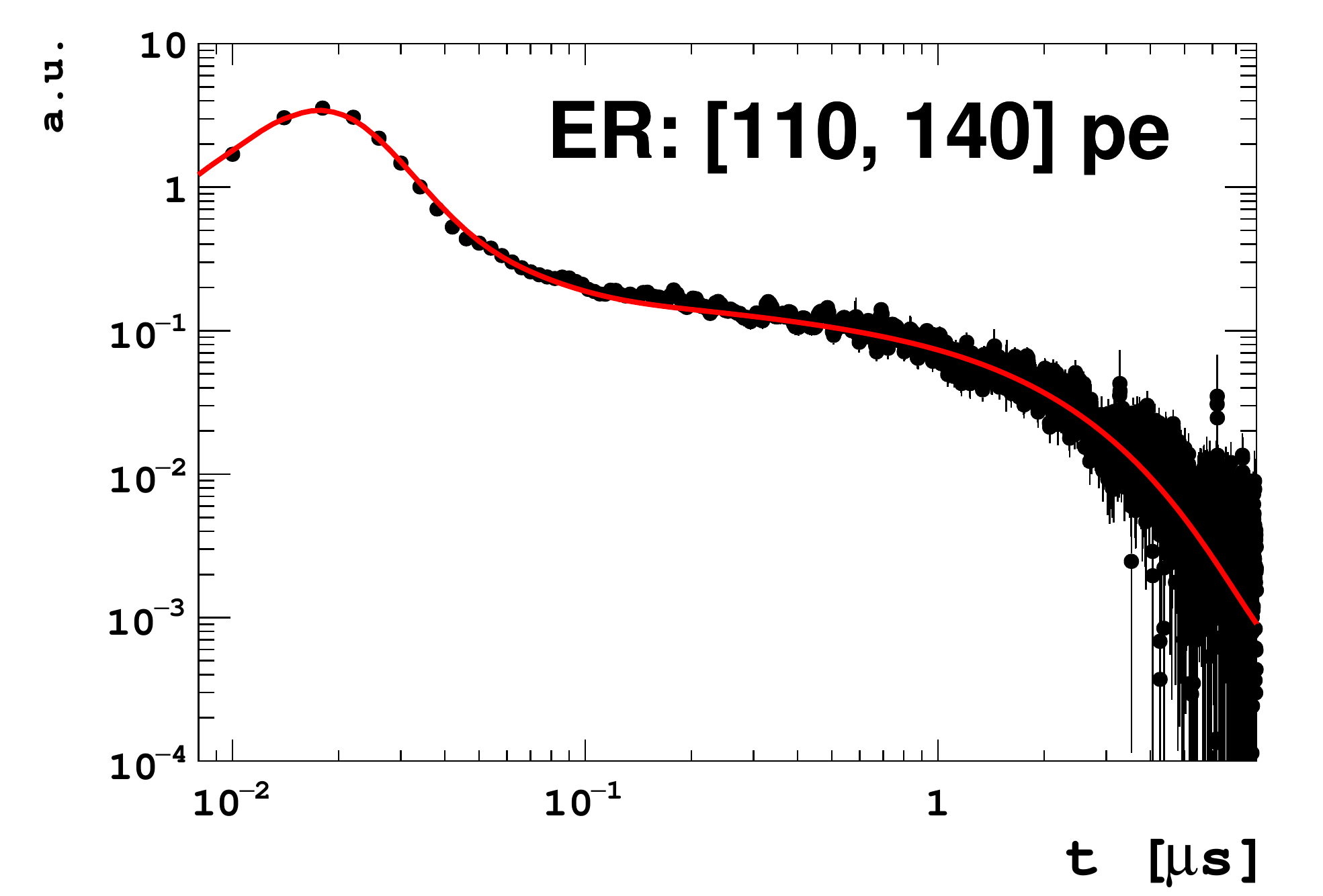} 
\includegraphics[width=0.31\textwidth]{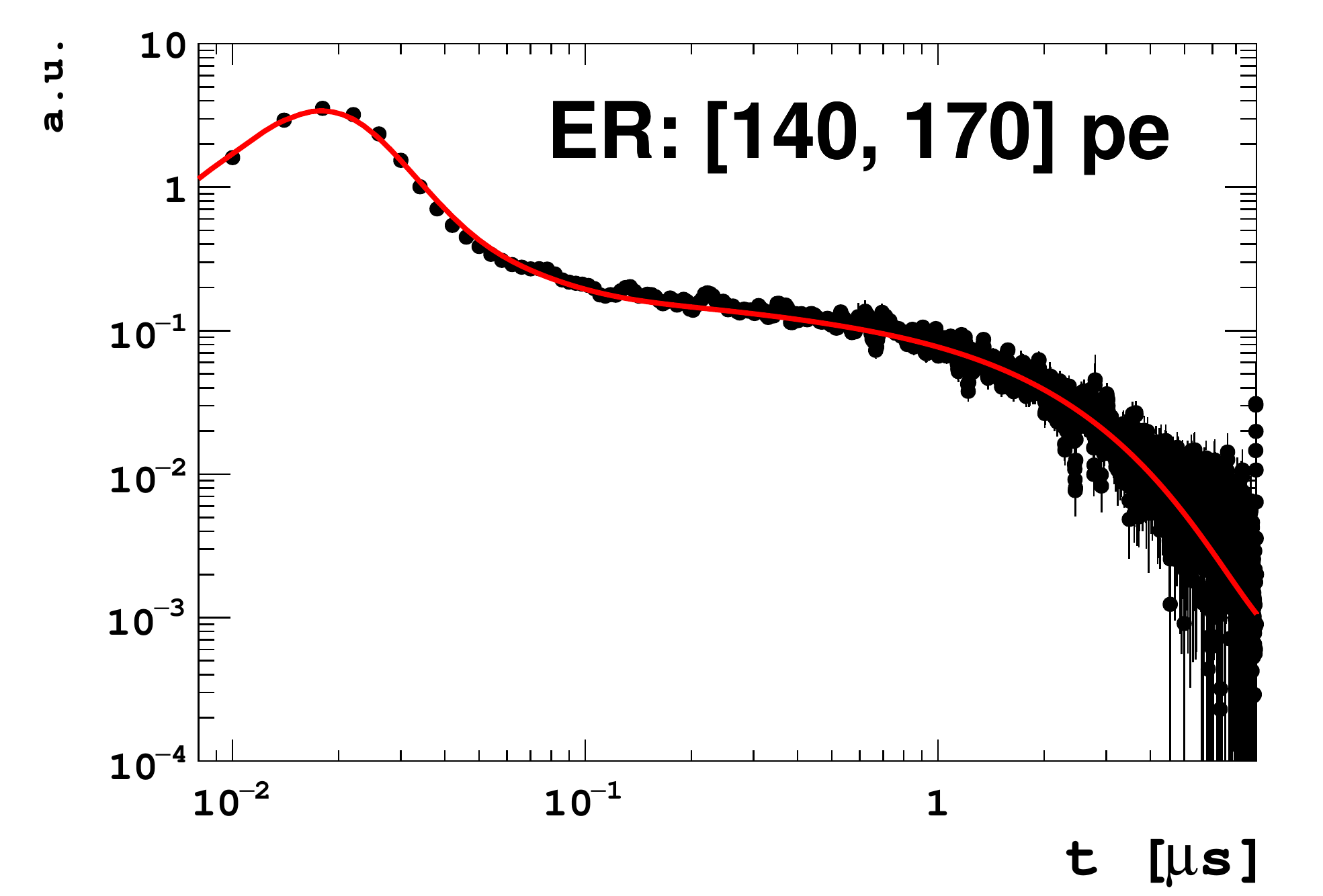}
\includegraphics[width=0.31\textwidth]{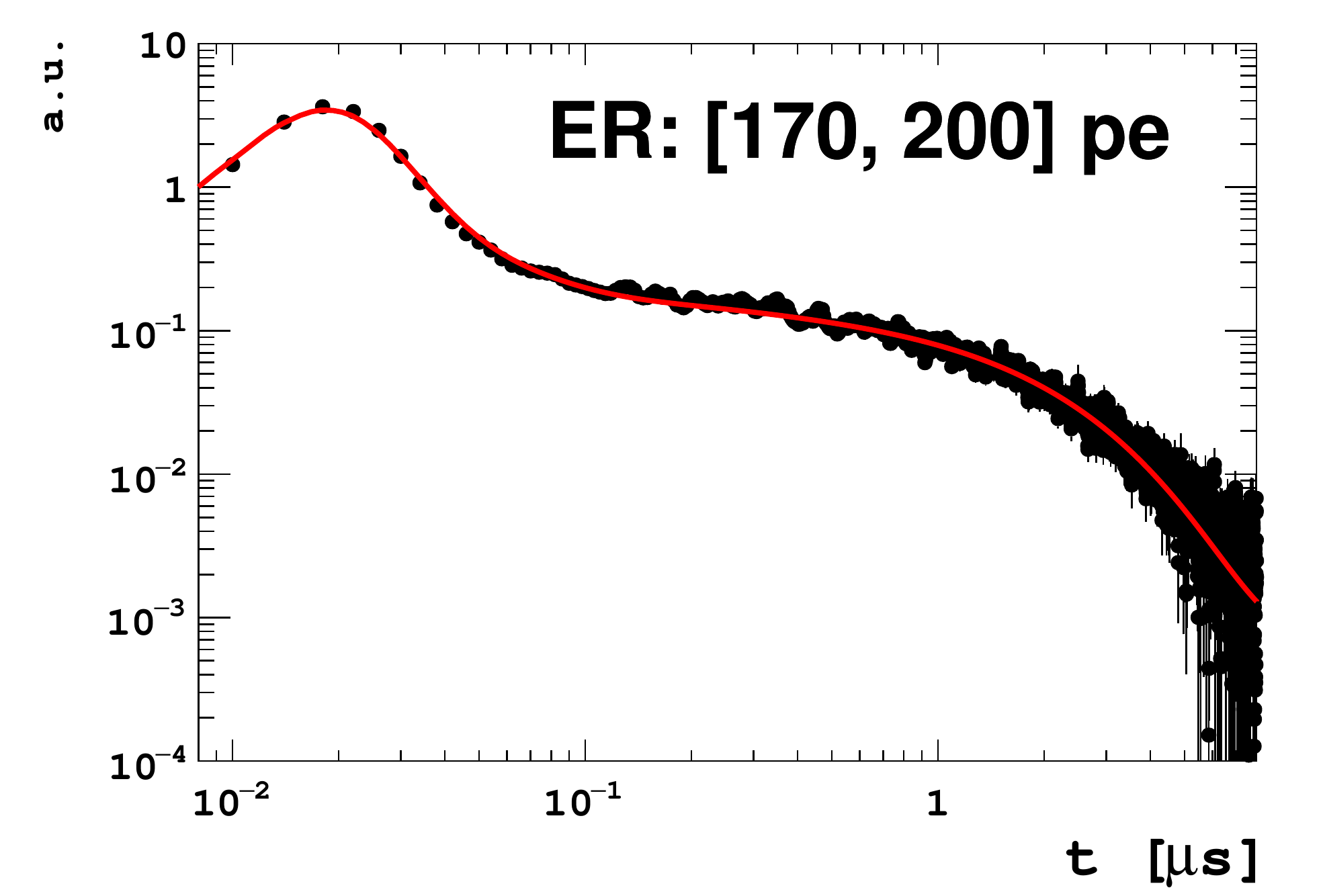}
\includegraphics[width=0.31\textwidth]{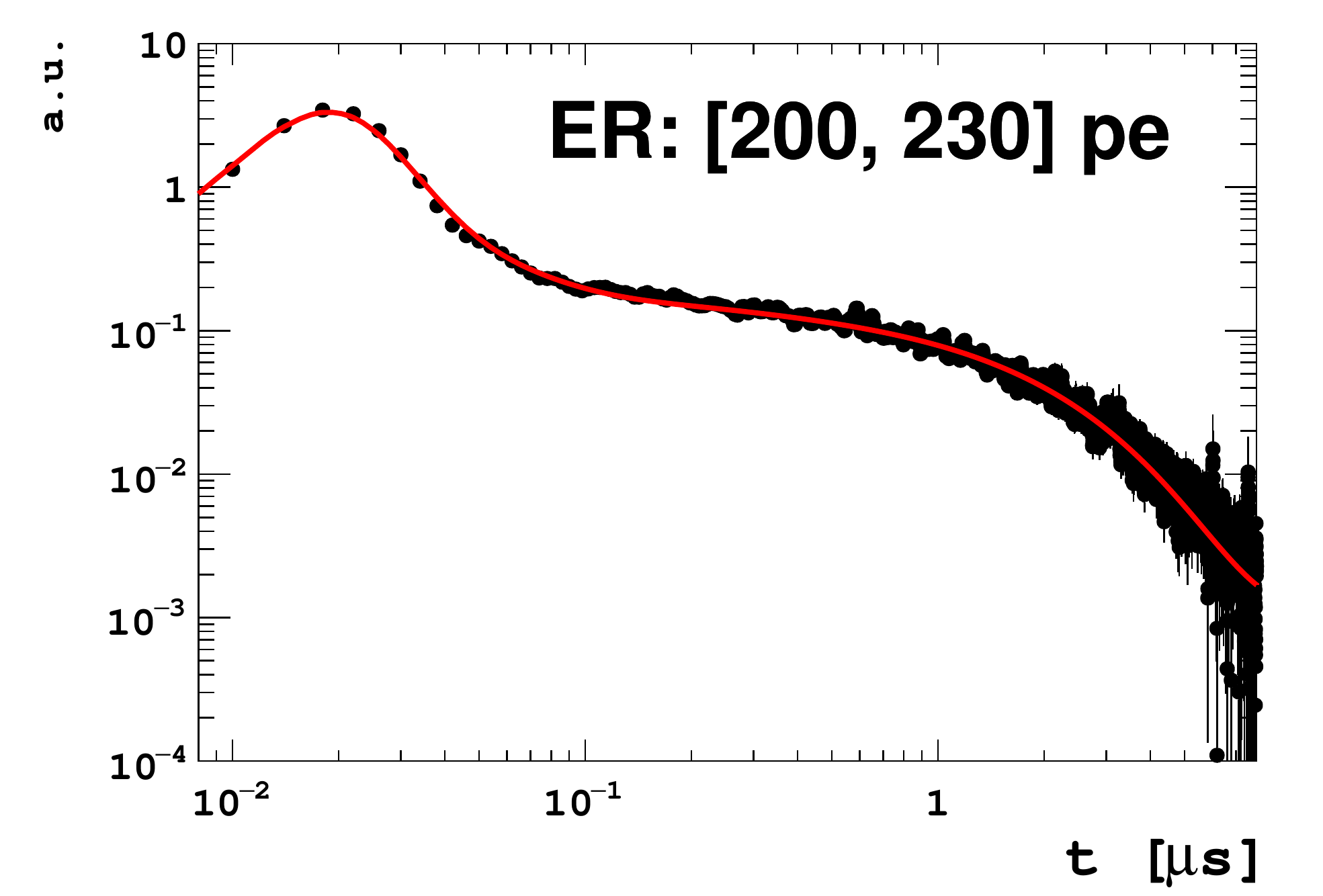}
\includegraphics[width=0.31\textwidth]{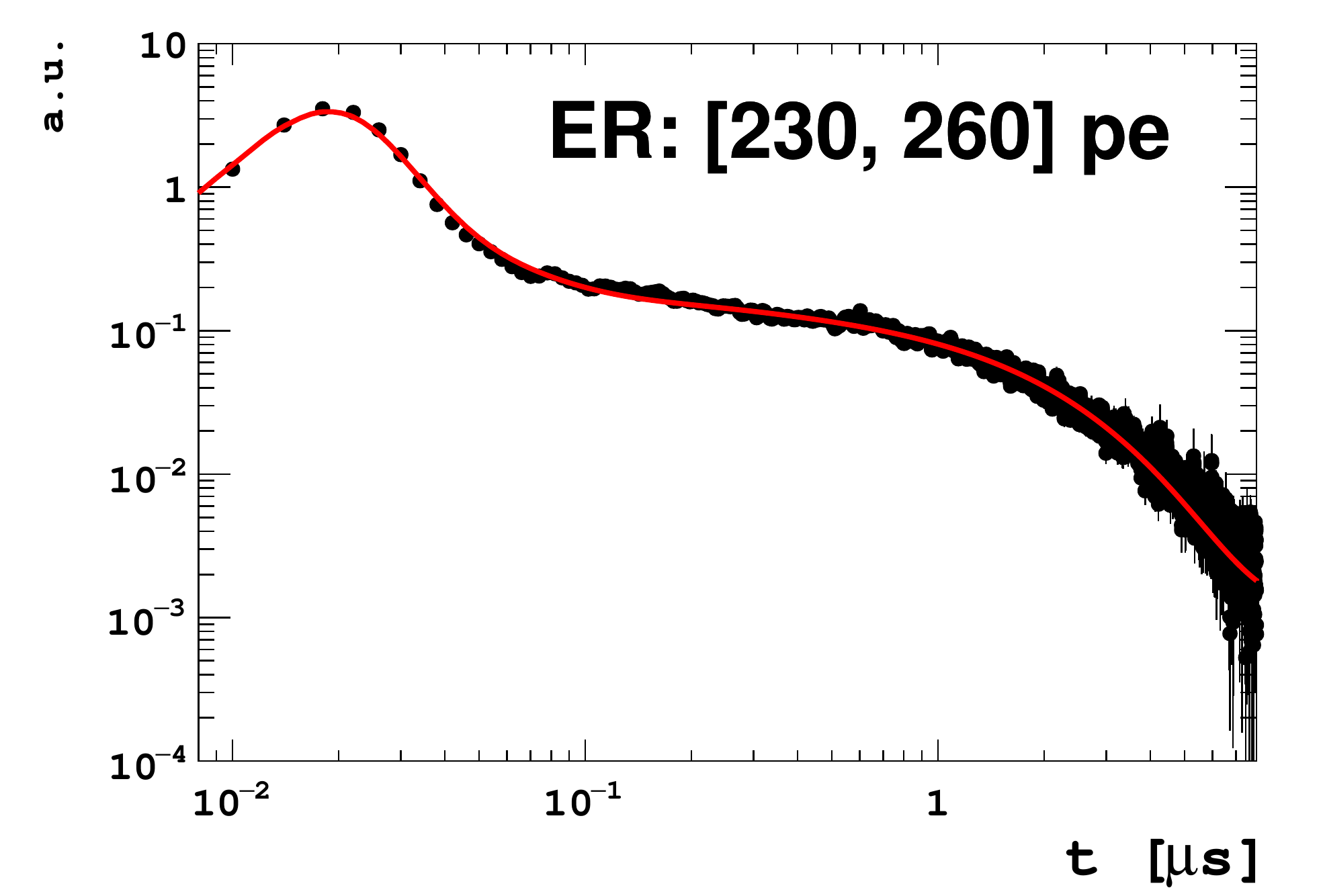}
\includegraphics[width=0.31\textwidth]{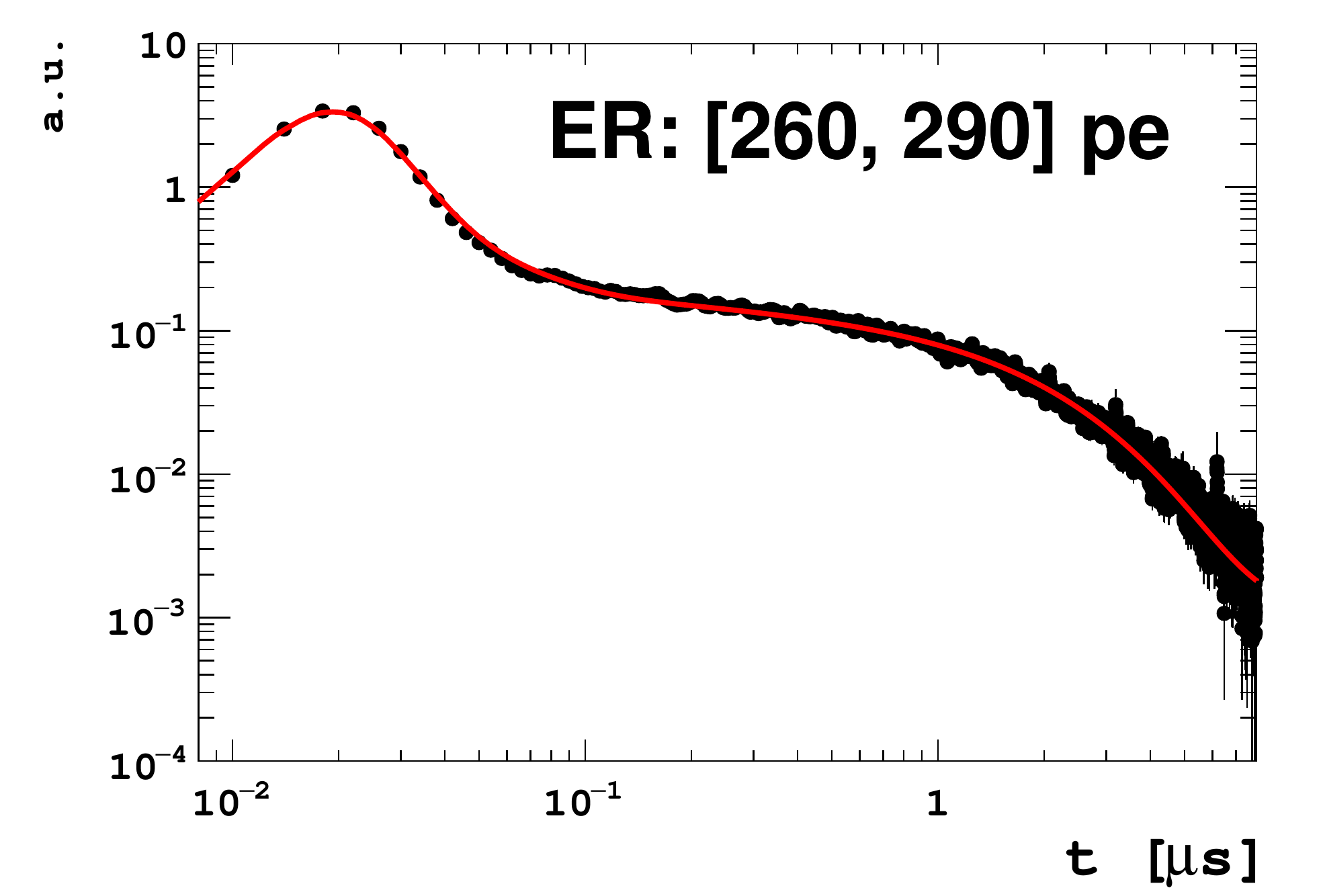}
\includegraphics[width=0.31\textwidth]{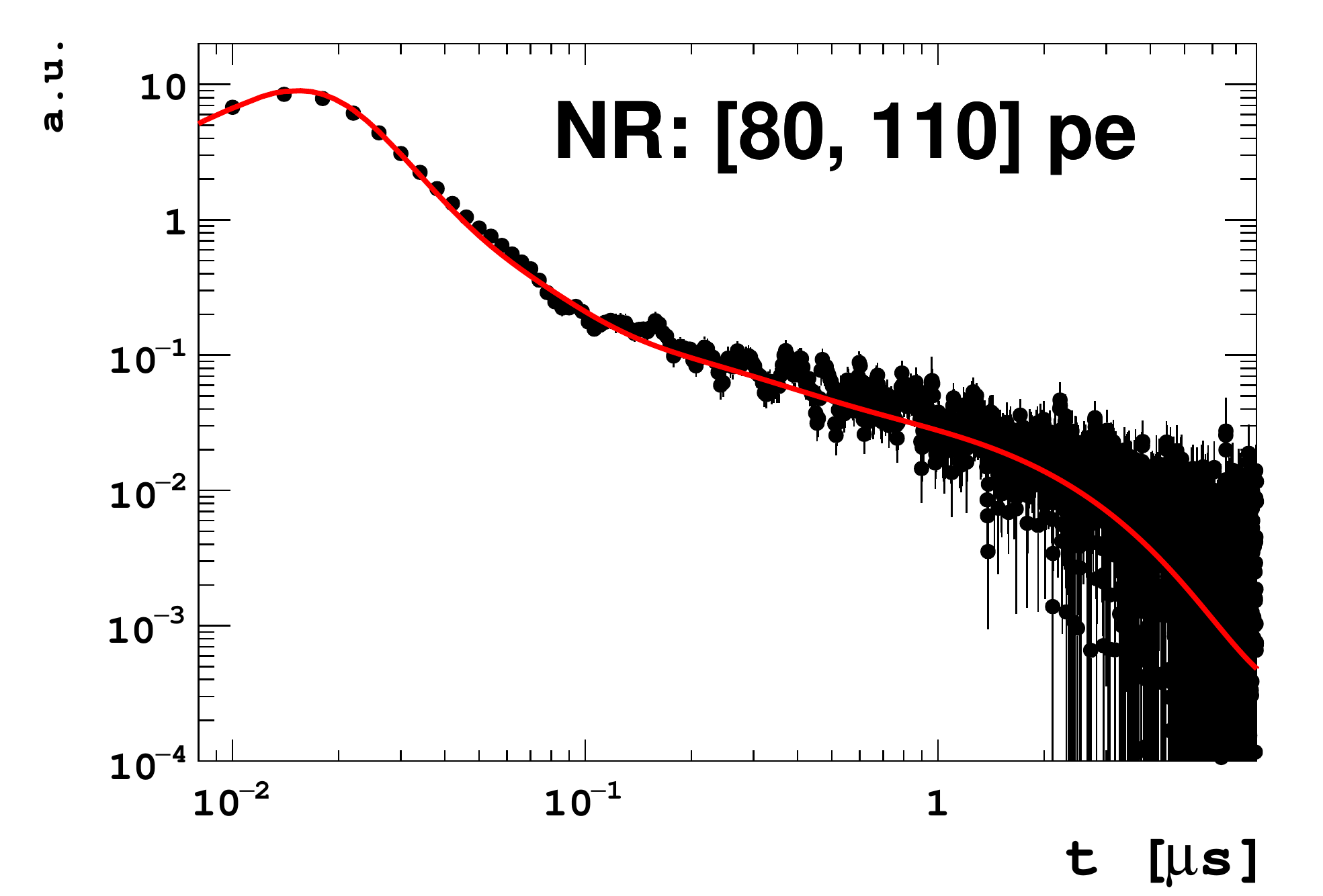}
\includegraphics[width=0.31\textwidth]{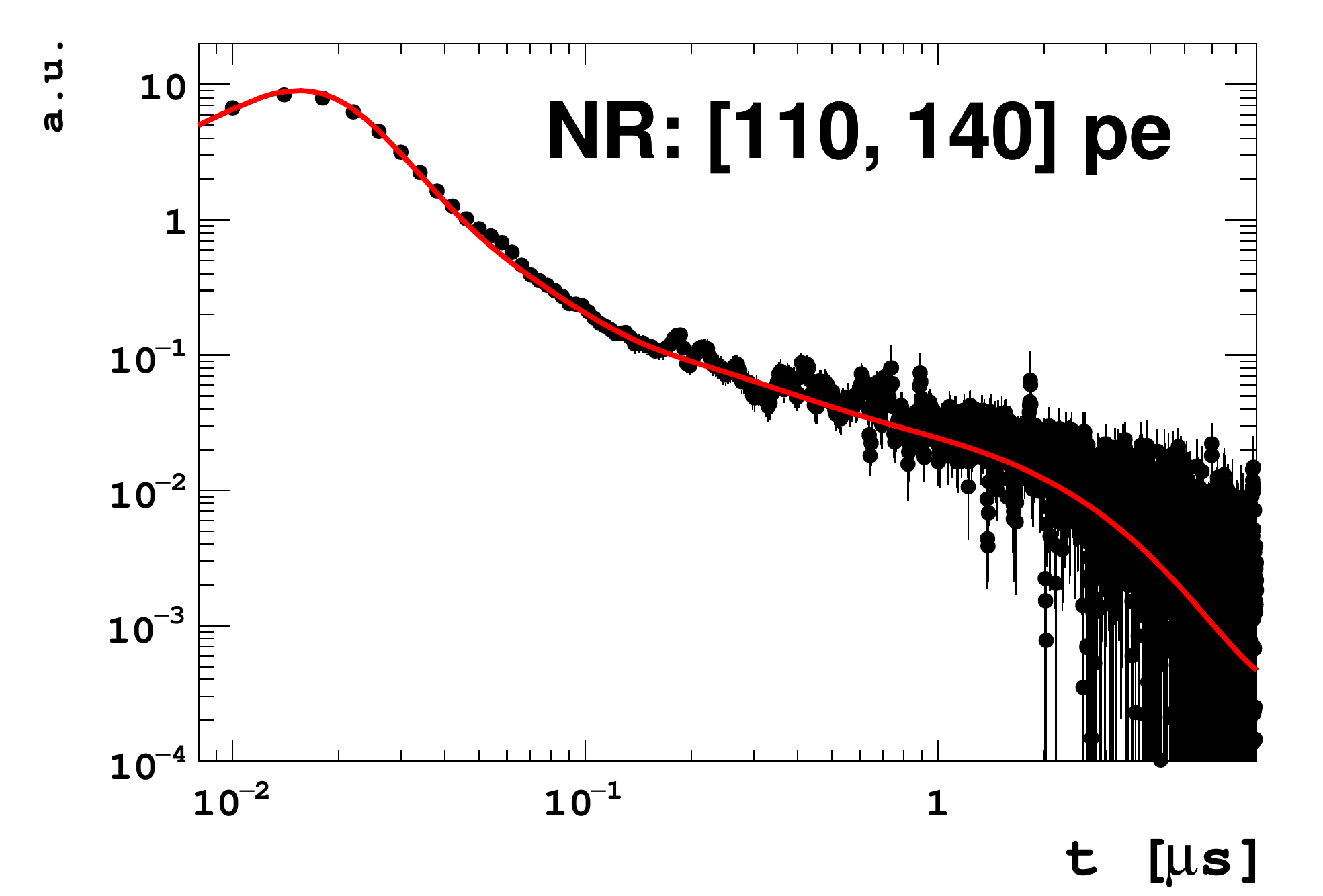}
\includegraphics[width=0.31\textwidth]{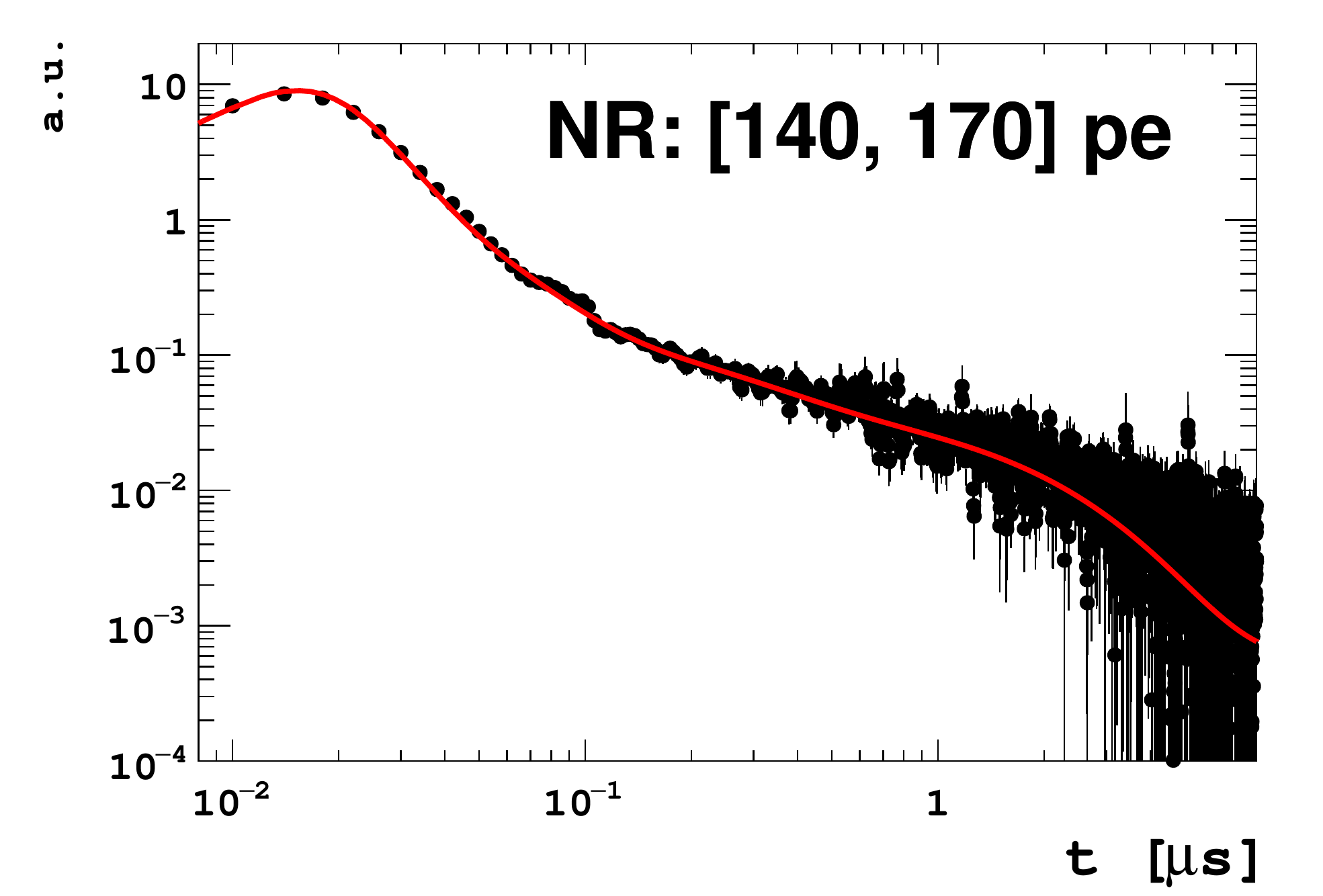}
\includegraphics[width=0.31\textwidth]{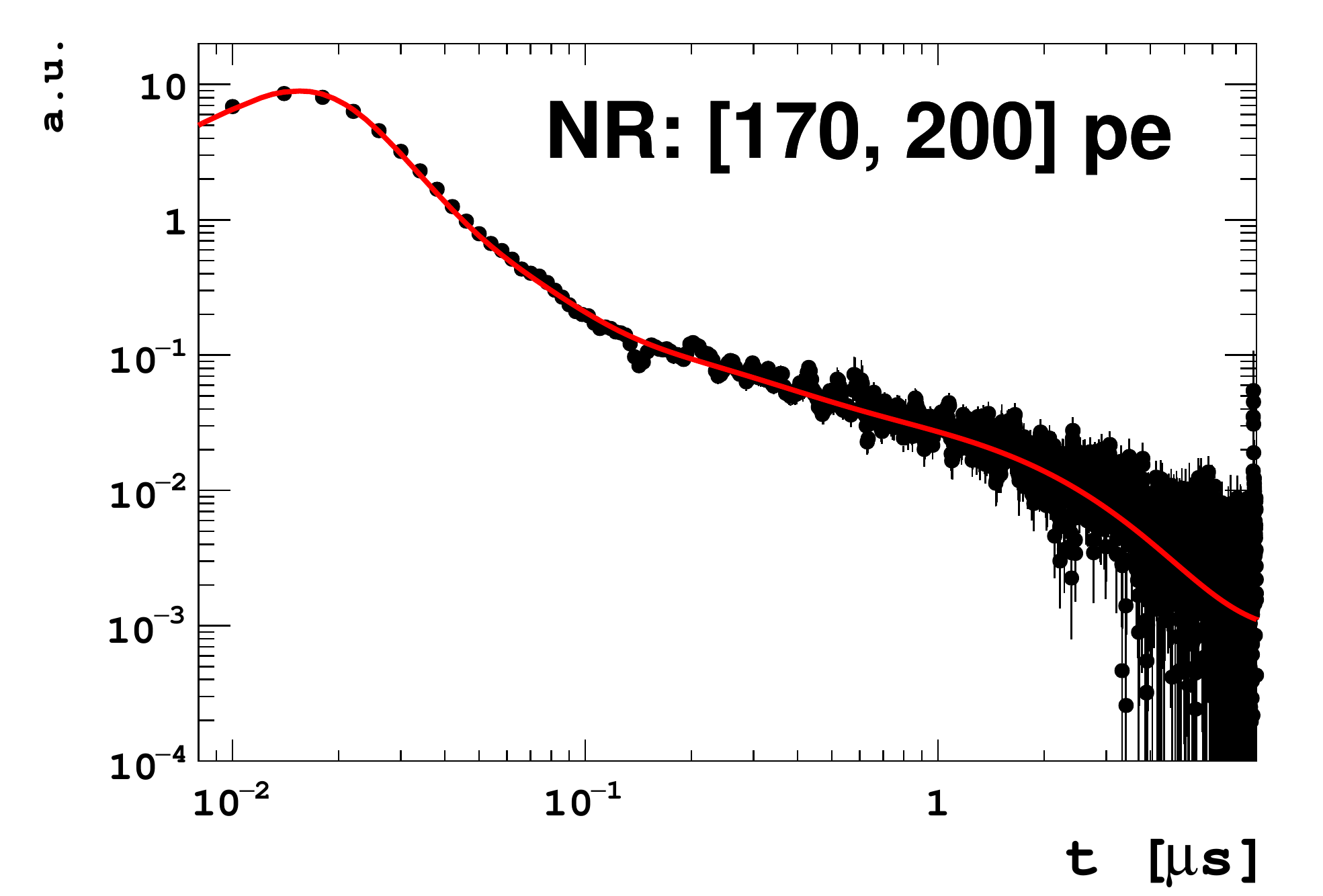}
\includegraphics[width=0.31\textwidth]{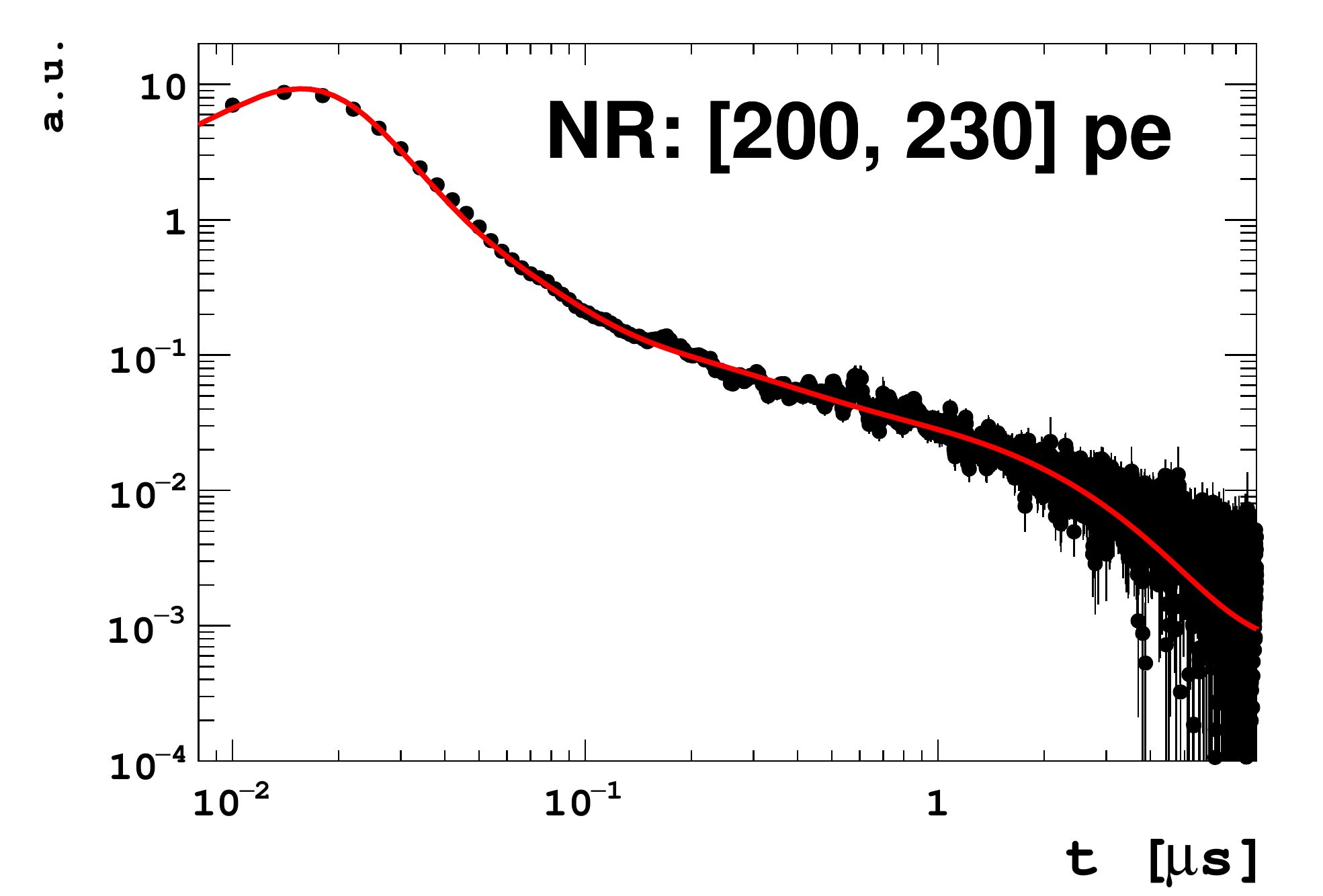}
\includegraphics[width=0.31\textwidth]{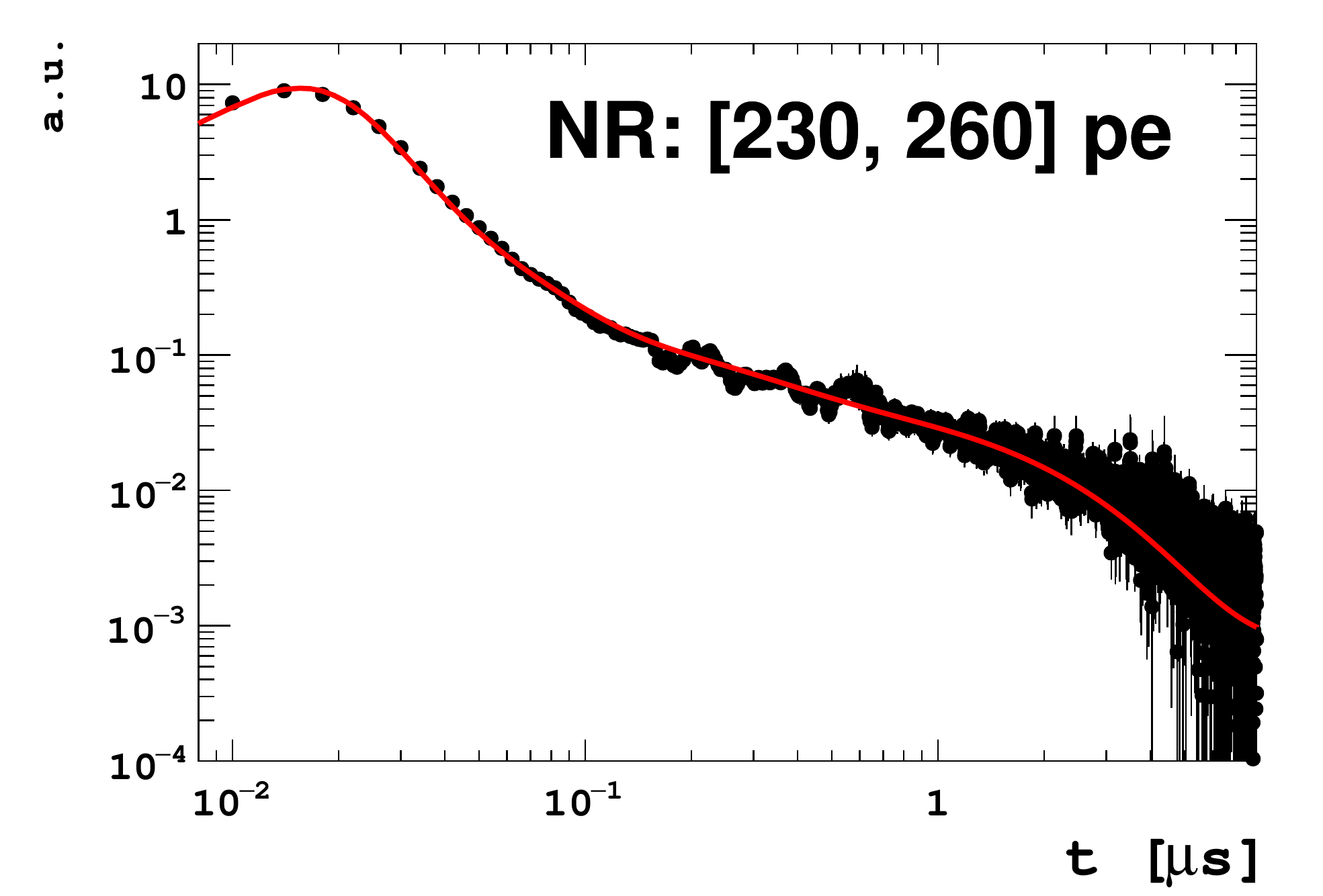}
\includegraphics[width=0.31\textwidth]{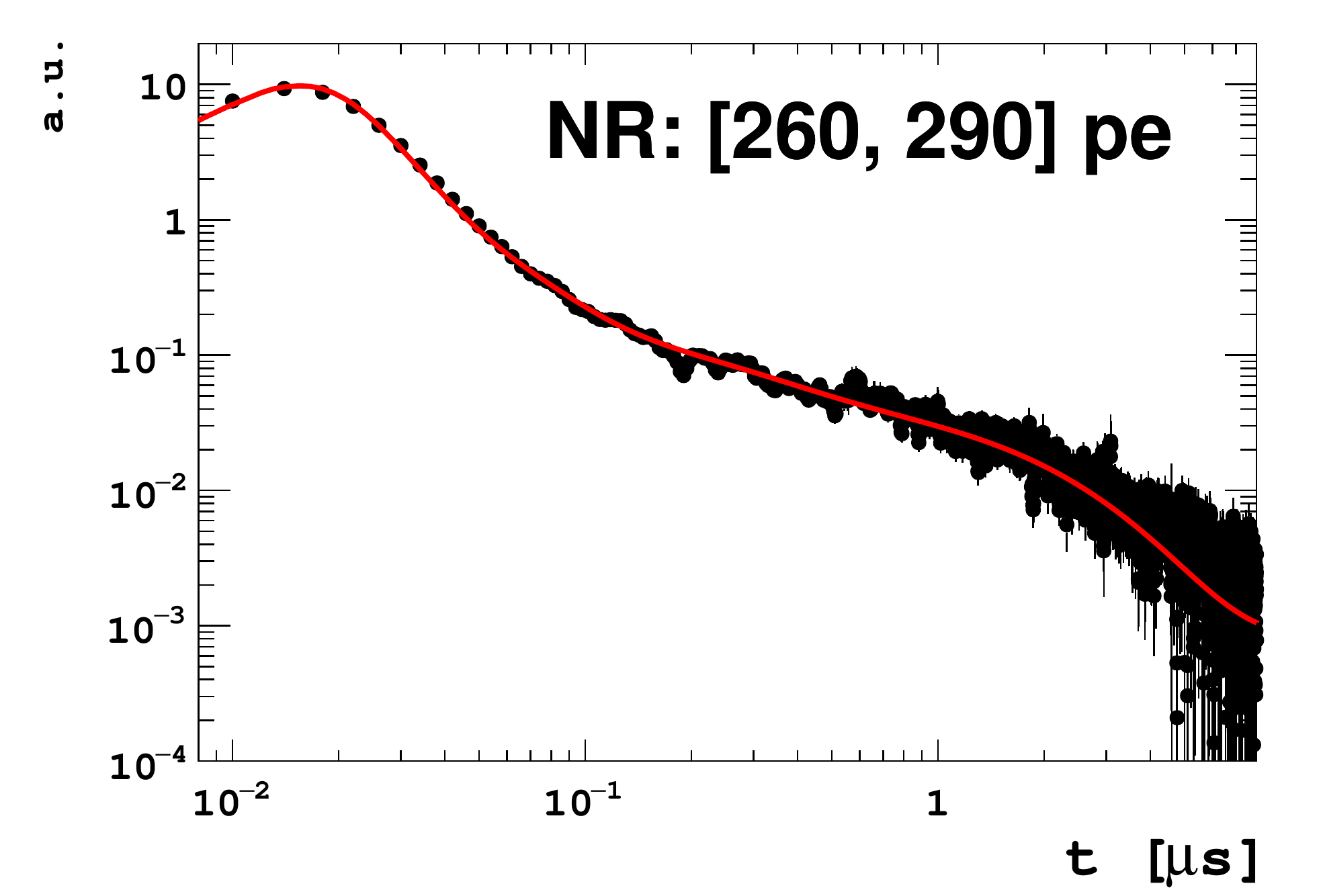}
\caption{Example of simultaneous fit of the 14 waveforms (black dots) for ERs   and NRs, in the 7 30-pe bins from 80 to 290\,pe, at 500\,V/cm. The red lines represent the fitting model.}
\label{fig:allWFs0}
\end{figure}

\subsection{Analysis and results}

\begin{table}[h]
\centering
\begin{tabular}{c | c | c | c | c | c }
\hline
TPB  & $\tau_t$  & $p_1$ & $\tau_1$ & $p_2$  & $\tau_2$ \\
\# & [ns] & [\%] & [ns] & [\%] & [ns] \\
\hline
0 & $1319\pm87$ & - & - & - & -   \\
1 &$1420\pm91$ & $14.7\pm0.6$ & $83\pm5$ & - & -   \\
2 &$1438\pm93$ & $14.5\pm1.1$ & $32\pm6$ & $9.1\pm0.9$ & $177\pm45$    \\
\hline
\end{tabular}
\caption{LAr scintillation slow component ($\tau_t$) and TPB delayed emission  time constants and probabilities ($\tau_j$ and $p_i$ with j=\{1,2\}) from the ARIS data fit assuming models with 0, 1, and 2 TPB components. }
\label{tab:tpb_results}
\end{table}

Each dataset is defined for a given electric field and consists of 14  waveforms, as already mentioned each averaged over about 10$^3$~events, corresponding to 7 energy ranges for each NR/ER sample. The free parameters associated to each waveform are $p_s$, $A$, $t_0$, and $C$.  Scintillation times, as well as  TPB and  detector parameters,  are  constrained  among all waveforms that make up a dataset. The total number of free parameters is 59 in addition to those associated to the TPB delayed emission. As the scope of this analysis is to probe and characterize the TPB fluorescence, the fit procedure was repeated assuming zero, one, and two TPB components.

At first, we tested a  systematic effect  potentially arising  from  the degeneracy between the scintillation fast component, $\tau_s$, and    the detector timing resolution, $\sigma$ (best fit yields $\sim$9 and $\sim$5\,ns for $\tau_s$ and $\sigma$, respectively). Since we noticed a non-negligible anti-correlation between the two parameters,  the fitting procedure was  repeated fixing $\tau_s$ to values in the  $[$5,12$]$\,ns range.  No significant  variation of the best fit parameters of interest, i.e., $p_s$, $\tau_s$, and TPB parameters, was observed. 

An example of fit with eq.~\ref{eqn:model2} of  the 14 waveforms acquired at 500~V/cm  is shown in Figure~\ref{fig:allWFs0}, assuming a single TPB delayed component.  The results of the fits for the three TPB configurations are quoted in Table~\ref{tab:tpb_results}. The associated errors are defined as the RMS of the mean values from the fits at different fields.

\begin{figure}[]
\captionsetup[subfigure]{labelformat=empty}
\centering
\includegraphics[width=0.55\columnwidth]{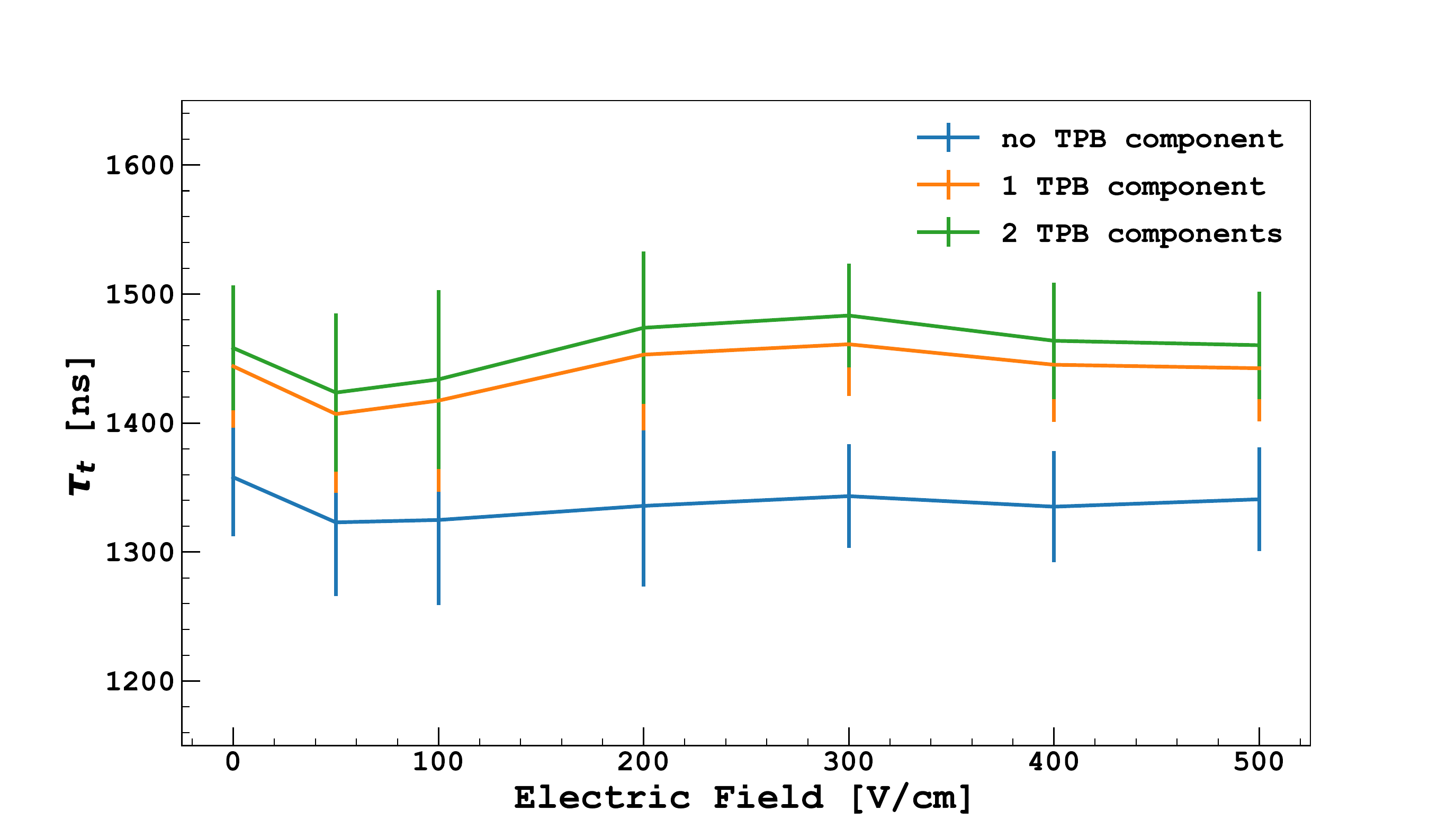} 
\caption{Slow scintillation decay time as a function of the electric field, assuming none, one, and two TPB components. }
\label{fig:tau_t}
\end{figure}

As a first test, we looked at the  dependence of the slow scintillation component on the electric field   suggested in ref. \cite{Aimard:2020qqa}, where  a 3~m$^3$ LAr TPC  observed, using cosmic ray events,  a $\sim$10\% reduction of $\tau_t$  by  varying the electric field  from 0 to 600~V/cm. In contrast,  this analysis did not highlight any significant deviation from a constant $\tau_t$ in any of the three TPB configurations, as shown in Figure~\ref{fig:tau_t}. %Furthermore, the $tau_s$ values returned by the three fits are all compatible within 1$\sigma$ with $\sim$1400~ns, as reported in Table~\ref{tab:tpb_results}. 

\begin{figure}[]
\captionsetup[subfigure]{labelformat=empty}
\centering
\includegraphics[width=0.45\columnwidth]{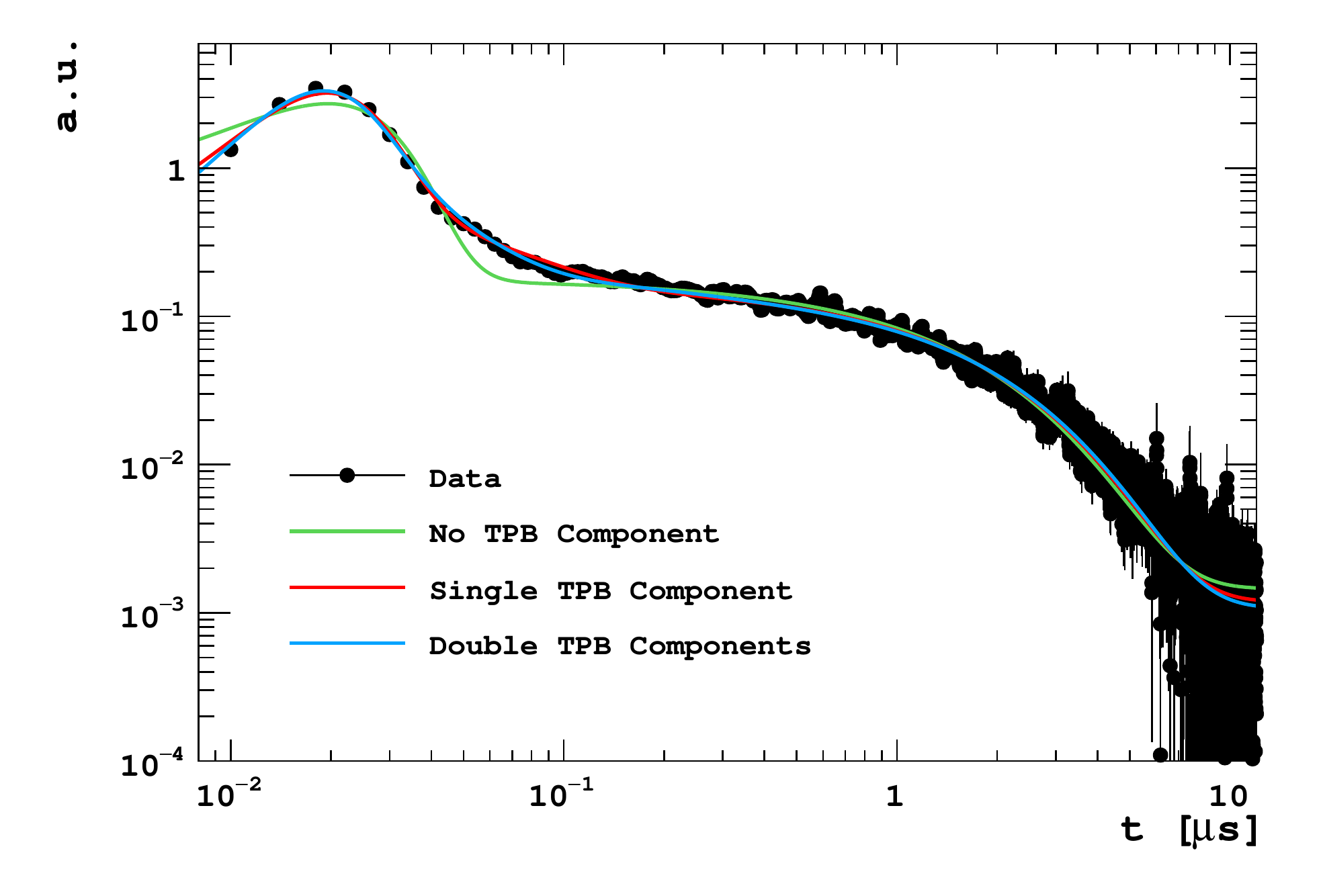} 
\includegraphics[width=0.45\columnwidth]{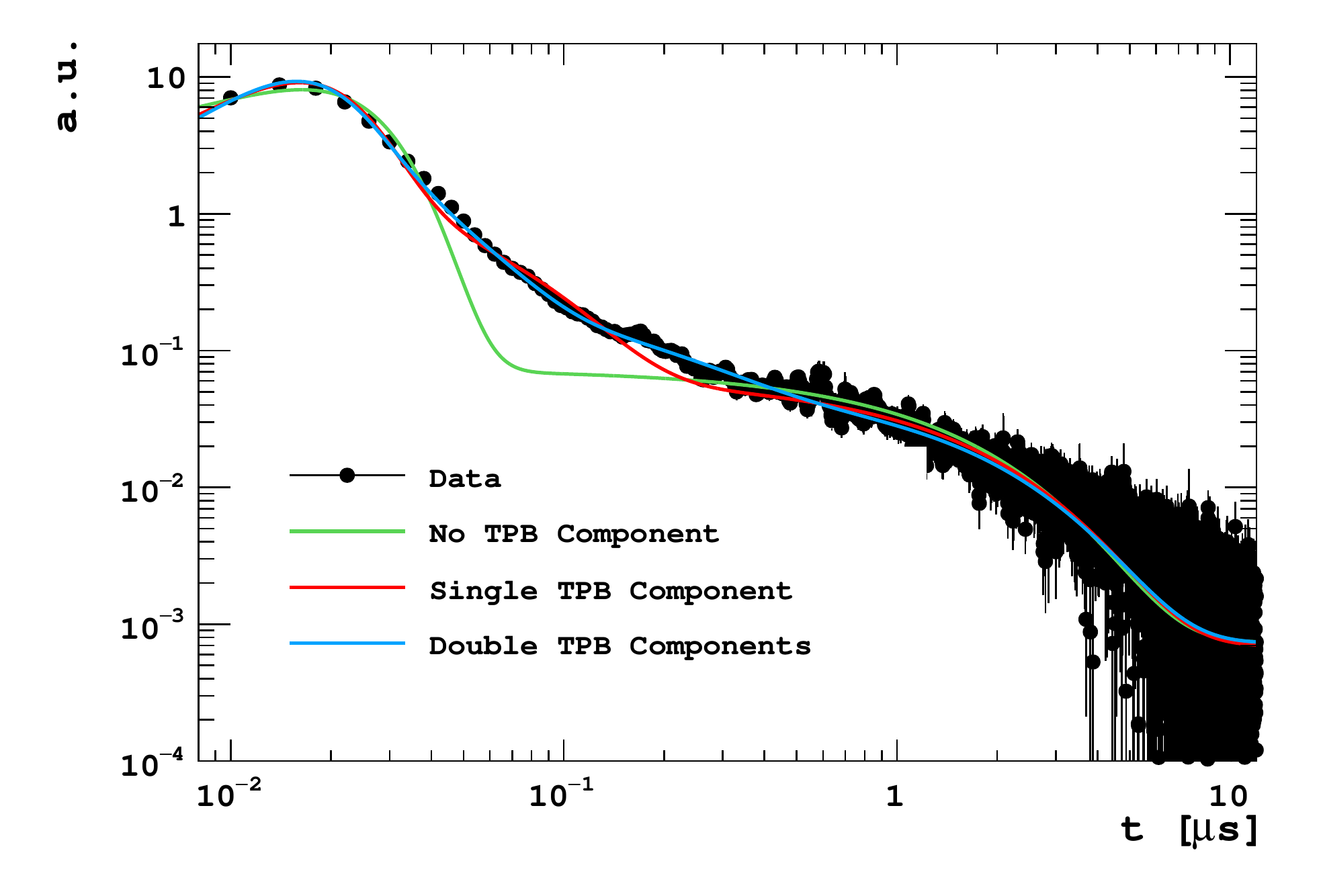} 
\caption{ER (left) and NR (right) averaged waveforms acquired with $500$\,V/cm drift field, for S1 in the [$200$, $230$]\,pe range. The fitting models assume none, one, and two TPB delayed components.}
\label{fig:wf500V}
\end{figure}

In Figure \ref{fig:wf500V} we show two examples of averaged waveforms taken at 500\,V/cm, fitted assuming none, one, and two TPB components. It can be noted that the NR spectrum cannot be modelled without at least one TPB component. This was confirmed by a null-hypothesis test, where the null-hypothesis corresponds to no TPB components (or $p_1$=0),  with respect to the one-component model (or $p_1$$>$0). The null-hypothesis was rejected in 98.5\% of  the fitted waveforms with   $\Delta\chi^2$  exceeding the 99\% C.L. equivalent threshold. The same testing procedure to verify whether the data preferred two versus one TPB components (here assumed as the null-hypothesis) did not produce any conclusive result, as the $\Delta\chi^2$ rejected the null-hypothesis with 99\% C.L. equivalent threshold for only 54.2\% of the fitted waveforms.  From this analysis, we conclude that the model with one TPB fluorescent component, with 83$\pm$5~ns decay time and 14.7$\pm$0.6\% probability,  is sufficient to  reproduce ARIS data.  This result is of the same order of magnitude as the dominant component measured in ref.  \cite{Segreto:2014aia} (49$\pm$1\,ns and 30$\pm$1\%) although not  compatible within the  uncertainties. It is worth noting, however, that Segreto's measurement was done in vacuum at room temperature, very different from the one presented in this paper performed in LAr.  In addition, the  acquisition gate used with ARIS, 7\,$\mu$s, is not sufficiently long to detect the  long TPB decay component  ($\sim$3.5\,$\mu$s with $\sim$8\% amplitude) reported in ref.~\cite{Segreto:2014aia}.

%% file: psd.tex
% !TEX root = paperARIS2.tex

The uncertainty of the the singlet-to-triple ratio, measured with the waveform fit described in the previous section,  is too large, because of the large number of free parameters, to infer its dependence on the electric field.  To overcome this problem, we implemented an effective description of the PSD estimator ($f_p$) distribution, which allows  to minimize the number of parameters and more accurately extract the $f_p$ dependence  on the electric field. $f_p$ describes the ratio of two correlated normal random variables. In the PSD context, they correspond to the number of prompt photelectrons, $n_p$, and to S1, so that 
\begin{equation}
w = \frac{n_p}{\text{S1}} =  \frac{n_p}{n_p+n_l},
\label{eqn:w}
\end{equation}
\noindent where $n_l$ is the number of photoelectrons detected in the late component. In  experiments like DarkSide-50 and DEAP-3600, the distribution of $w$, $f_p(w)$ is built  by selecting a narrow S1 range.

The here-proposed  model  considers an infinitely small interval in the neighborhood of S1=S1$_0$, so that
\begin{equation}
n_p + n_l = \text{S1}_0.
\label{eqn:constraint}
\end{equation}

The $w$ observable is then the ratio between a random variable, $n_p$, constrained  by $n_p \leq \text{S1}_0$, and  S1$_0$, and its variance is
\begin{equation}
\sigma_w^2 = \left(\frac{\partial w}{\partial n_p}\right)^{2} \sigma_p^2 + \left(\frac{\partial w}{\partial n_l}\right)^{2} \sigma_l^2  + 2\,\frac{\partial w}{\partial n_p} \frac{\partial w}{\partial n_l} \sigma_{pl},
\label{eqn:var}
\end{equation}
\noindent where  $\sigma_p$ and $\sigma_l$ are the prompt and late component resolutions, and  $\sigma_{pl}$ is the covariance term. 

The main assumption at the basis of this model is that any physical effect playing a role in the definition of $n_p$ and  $n_l$ can be modelled with Poisson and Binomial statistics. The number of photoelectrons emitted by the LAr scintillation can be, in fact, well approximated by a Poisson distribution, as shown in reference \cite{Agnes:2017grb}. In addition, any effect related to photon emission, propagation, and detection is associated to a given probability to contribute to either the prompt or late component, and thus can be regarded in this context, as a Bernoulli process. Their composition gives origin to a Binomial distribution. Additionally, the correlated noise (e.g. afterpulses) contributes binomially to prompt/late signals.

Both Poisson and Binomial distributions can be approximated, at sufficiently large photon statistics, by Gaussian distributions, whose standard deviation is proportional to the square root of the number of photoelectrons. As their convolution is still Gaussian with the same dependence on the number of photoelectrons,  we introduce the following definition of prompt and late resolutions,
\begin{equation}
\sigma_i  =    k_i \times \sqrt{\text{S1}_0}
\label{eqn:var2}
\end{equation}
\noindent for $i=\{p,l\}$ and $k_i$ constant.

Following our assumption, $w$ is also Gaussian distributed as 
\begin{eqnarray}
f_p(w)  =  \frac{e^{ -\frac{(w - w_0)^2}{2\sigma^2(w)}}}{\sqrt{2\pi\sigma^2(w)}},
\label{eqn:f90}
\end{eqnarray}
\noindent where $w_0$ is the most probable value  of $w$,  and $\sigma_w(w)$ is the standard deviation, 
\begin{equation}
\sigma^2(w) = (1 - w)^2  k_p^2   + w^2  k_l^2  + 2  w (1-w) k_p \, k_l,
\label{eqn:sigma}
\end{equation}
\noindent derived from  eqs. \ref{eqn:var}, \ref{eqn:var2}, and taking into account the full anti-correlation from eq. \ref{eqn:constraint} between $n_p$ and $n_l$.

\begin{figure}[t]
\centering
\includegraphics[width=0.45\columnwidth]{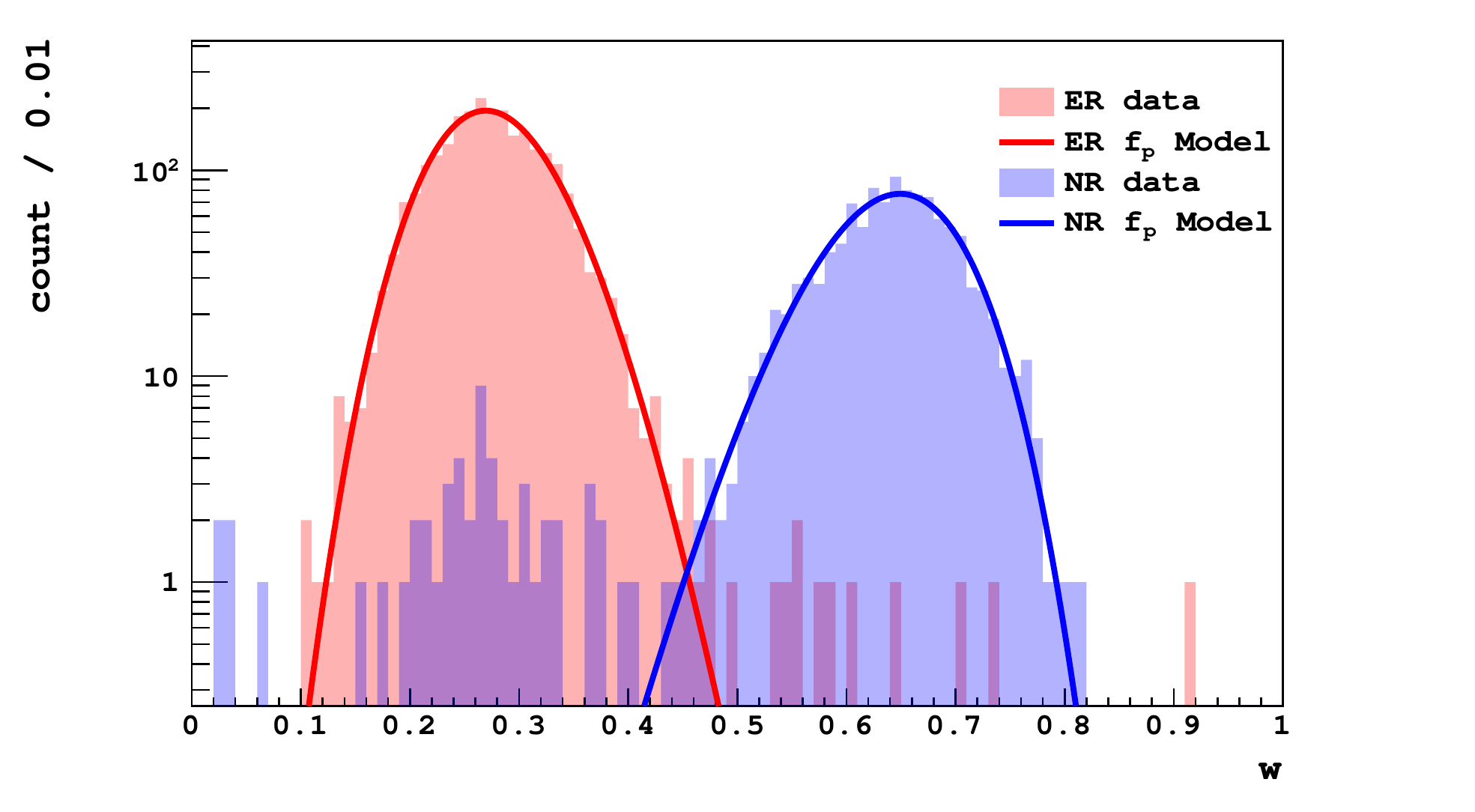}
\includegraphics[width=0.45\columnwidth]{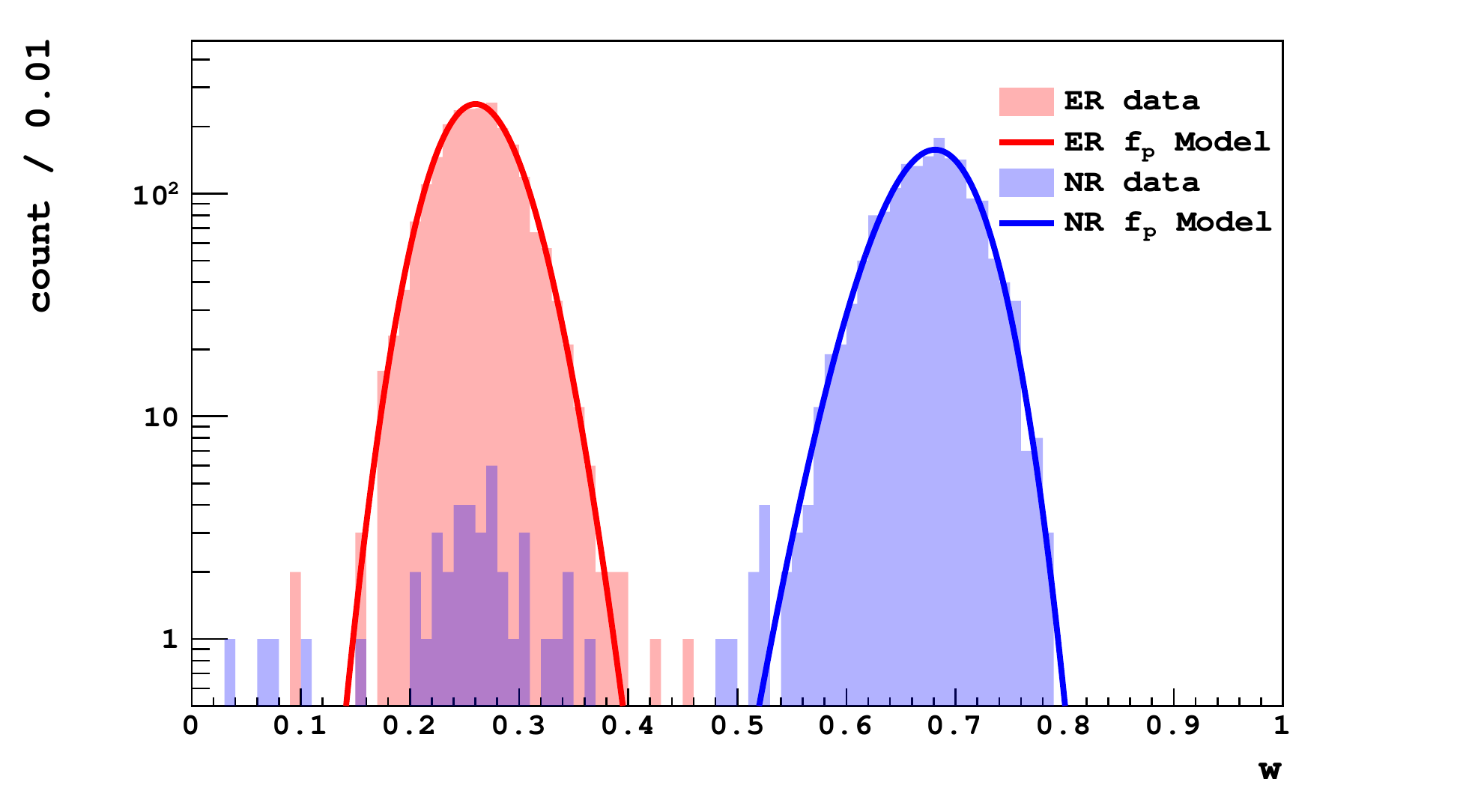}
\caption{ Fit of ARIS ER and NR $f_p$ distributions with eq.~\ref{eqn:f90} for S1=100~pe (left) and S1=200~pe (right).   }
\label{fig:fit_f90}
\end{figure}

Eq. \ref{eqn:f90}, successfully tested on Monte Carlo samples, was used to fit ARIS data with $f_p$ defined with a 90~ns prompt window and $f_p$ distributions selected with 5-pe S1 bins for each electric field. NR events are selected using the beam-TPC time coincidence, as described in Section~\ref{sec:setup}. In order to maximize the ER event statistics, we analyzed calibration data acquired with  a $^{133}$Ba source. Examples of fit for ERs and NRs  are shown in Figure~\ref{fig:fit_f90}.

\begin{figure}[t]
%\captionsetup[subfigure]{labelformat=empty}
\centering
\includegraphics[width=0.7\columnwidth]{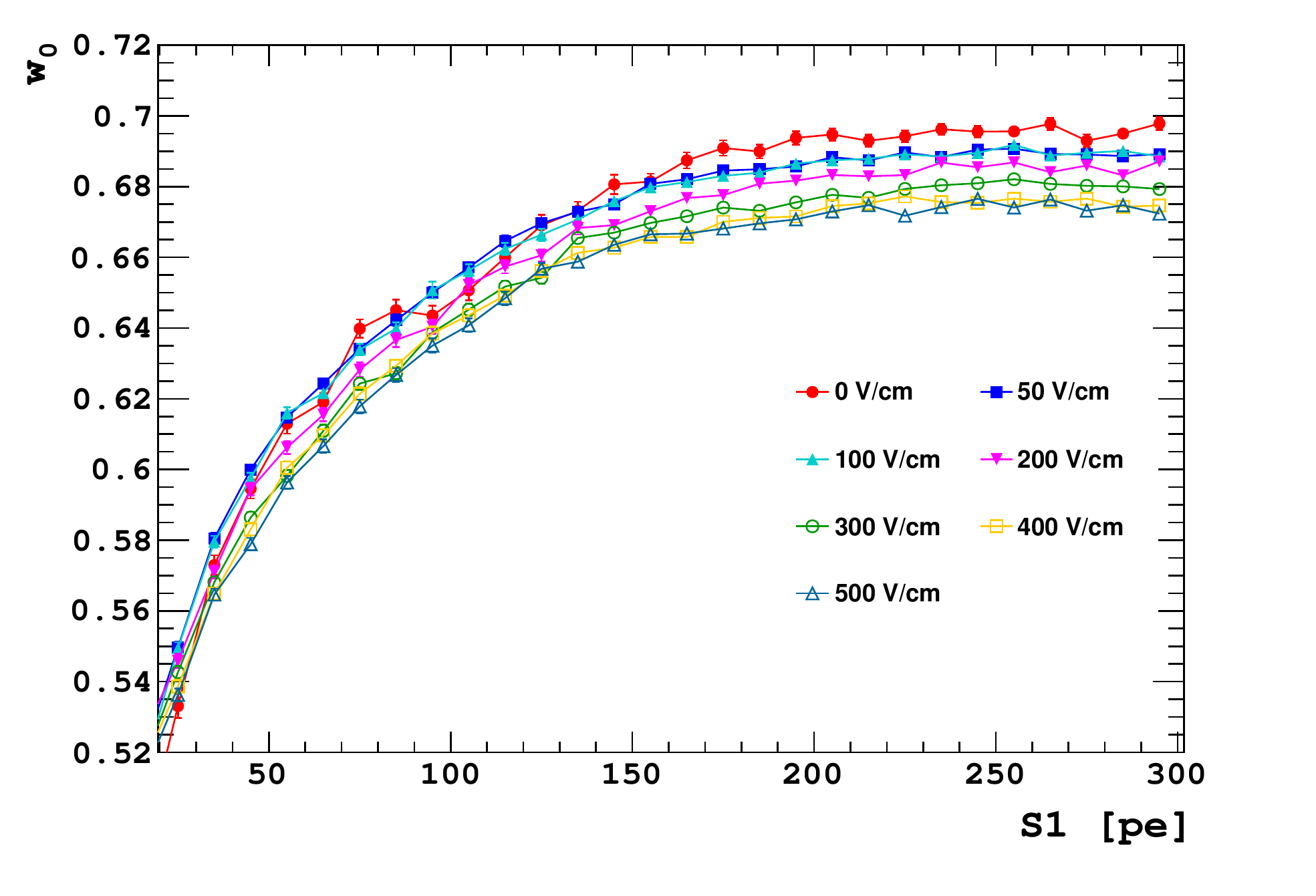}
\includegraphics[width=0.7\columnwidth]{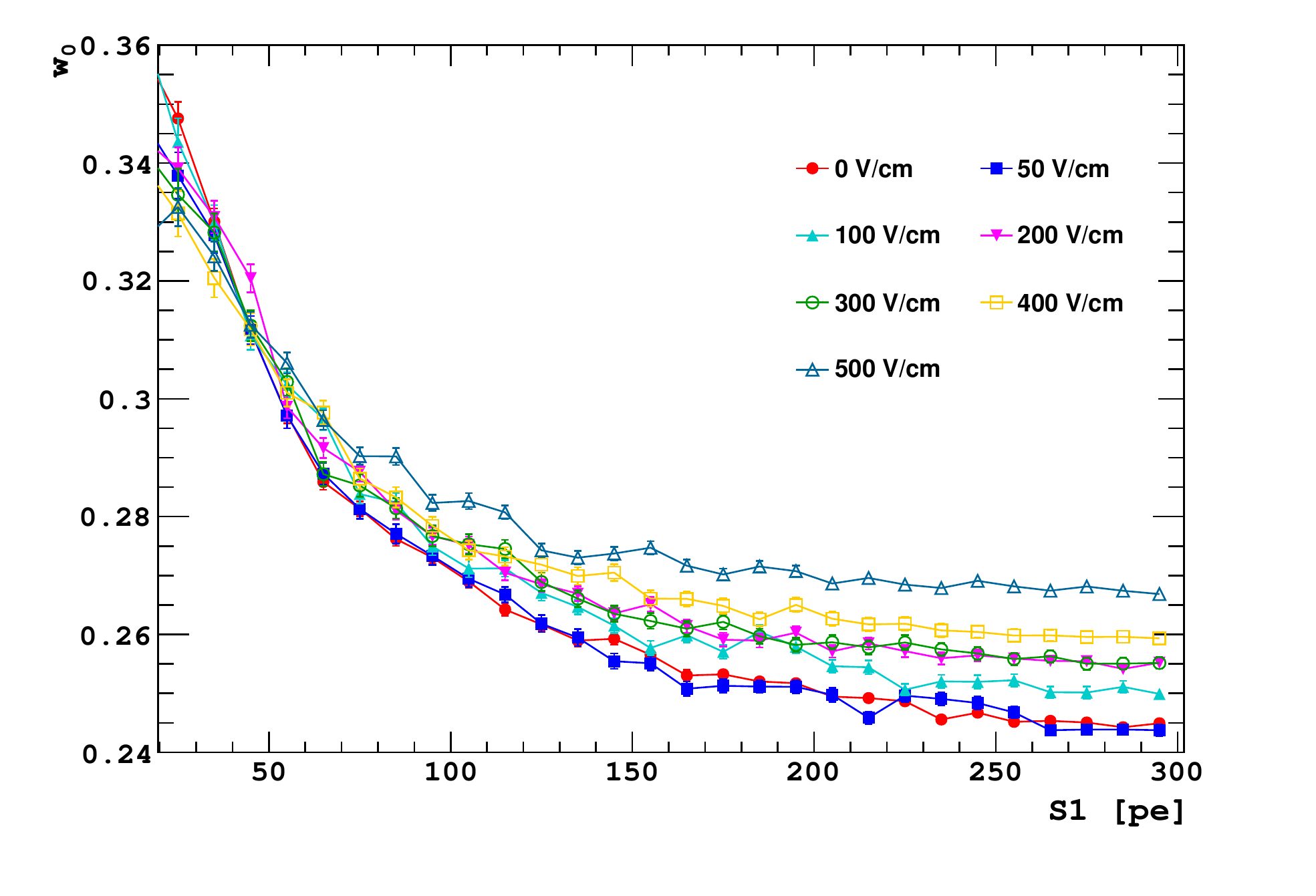}

%\subfloat[][]{\includegraphics[width=\columnwidth]{./fig/w0_s1_ER.pdf}\label{fig:w0_s1_er}} \\
%\subfloat[][]{\includegraphics[width=\columnwidth]{./fig/w0_s1_NR.pdf}\label{fig:w0_s1_nr}} \\

\caption{NR (top) and ER (bottom) $w_0$ dependence on  S1 and on the electric drift field.}
\label{fig:allw0s}
\end{figure}

The dependence of $w_0$ on S1 and on the drift field for ERs and NRs are shown in Figure \ref{fig:allw0s}. The NR $w_0$ dependence on the electric field confirms what was already observed by the SCENE~\cite{Cao_2015} experiment, \textit{i.e.}   $w_0$ decreases as the field strength increases.  In addition to this, we  observe for the first time the dependence of ER $w_0$ on the field, which behaves opposite to NRs, \textit{i.e.}  $w_0$ increases as the field strength decreases.  Since we did not  observed any dependence of the  scintillation triplet de-excitation time on the electric field, as discussed in the previous section, we assume that the electric field may  act on the singlet-to-triplet ratio. However, we cannot provide any explanation to support this observation, especially given the opposite dependence of ER and NR on the electric field. We therefore report this result, confident that it will stimulate interest in the atomic and nuclear physics communities to better understand the mechanisms underlying the interaction between LAr scintillation and the electric field. 

%\begin{figure}[h]
%\centering
%\includegraphics[width=\columnwidth]{./fig/w0_ene_NR.pdf}
%\caption{  $w_0^{NR}$  as a function of the recoil energy for different electric fields.
%}
%\label{fig:w0_ene}
%\end{figure}
%
%For completeness of information and for possible utility in future experiments, we also show in Figure~\ref{fig:w0_ene} the dependence of NR $w_0$ as a function of recoil energy, thanks to the accurate ARIS calibration of the NR energy scale~\cite{Agnes:2018mvl}. 